\documentclass[12pt,preprint]{aastex}
\notetoeditor{refer to: Fiamma Capitanio. Institute: IASF-INAF,Via Fosso del Cavaliere, 100, 00133 Rome ITALY. Tel: +390649934557. E-mail: fiamma.capitanio@rm.iasf.cnr.it}
\newcommand{\inst}{\altaffilmark}

\usepackage{longtable}
\usepackage{graphicx}
\usepackage{lscape}
\usepackage{natbib}
\shorttitle{BeppoSAX WFCs Survey}
\shortauthors{Capitanio F. et al.}
\begin{document}

\title{Finding persistent sources with the BeppoSAX/WFC: a in-depth analysis}

\author {F. Capitanio\inst{1}, A. J. Bird\inst{2}, M. Fiocchi\inst{1}, S. Scaringi\inst{3}, P. Ubertini\inst{1}}
 
\altaffiltext{1}{INAF IASF-Roma, Via Fosso del Cavaliere 100, 00033 Rome,
Italy}

\altaffiltext{2}{School of Physics and Astronomy, University of Southampton, Highfield, Southampton, SO17 1BJ, UK}
\altaffiltext{3}{Department of Astrophysics, Radboud University Nijmegen, PO Box 9010, 6500 GL Nijmegen, The Netherlands}


\begin{abstract}

During the operational life of the Italian/Dutch X-ray satellite (1996-2002), BeppoSAX, its two Wide Field Cameras performed observations that covered the full sky at different epochs. Although the majority of analysis performed on BeppoSAX WFC data concentrated on the detection of transient sources, we have now applied the same techniques developed for the {\it INTEGRAL}/IBIS survey to produce the same work with the BeppoSAX WFC data. This work represents the first unbiased source list compilation produced from the overall WFC data set optimised for faint persistent sources detection. This approach recovers 182 more sources compared to the previous WFC catalogue reported in \citet{Verrecchia}.  
The catalogue contains 404 sources detected between 3-17 keV, 10 of which are yet to be seen by the new generation of telescopes.
\end{abstract}


\section{Introduction}

The two Wide Field Cameras (WFCs) \citep{Jager} on board the BeppoSAX
satellite \citep{Boella}, were mounted 180 degrees away from each other and pointed perpendicular to the direction of the Narrow Field Instruments (NFI), hence looking at two different sky zones during each NFI pointing. In this way, over the 6 years operational life of BeppoSAX, the WFCs observations covered almost all the sky with at least one pointing  (typically 100 ks duration). This serendipitous observing strategy, during which the WFCs acted as secondary instruments, was driven by the approved Narrow Field Instrument observing programme. However, twice a year, for around 8\% of the total observing time of the satellite, the WFCs observed the Galactic bulge region as primary instruments (i.e. thanks to a pre-planned observing programme) collecting a total exposure on the Galactic Centre of 6 Ms during the operational life of the satellite.

The WFCs were coded mask instruments characterised by a large ($40\deg \times 40\deg$) field of view, a good angular resolution (few arcmin) and a pointing strategy that permitted all the sky to be observed during the satellite's operational life.  The operating principles are reported in detail in \citet{Jager}. The principal scientific objective of the WFCs on BeppoSAX was the study of the X-ray variability of the sky. In fact, through the serendipitous monitoring of large sky regions, the WFCs were able to detect a lot of transients X-ray events like GRBs and X-ray binary outbursts  \citep[see e.g.][and reference therein]{Piro}. Most of the time, the WFCs were used as triggers for follow-up studies with higher-sensitivity narrow-field instruments on BeppoSAX itself or on other platforms. The principal characteristics of the WFCs are briefly summarised in Table~\ref{charact}.

The goal of our study is to reanalyse the WFC data so as to obtain a static view of the sky averaged all over the six years of BeppoSAX operational life in order to search for faint persistent sources that remained hidden in the previous analyses because they were too faint to be detected with an adequate confidence level in a single observation. 

This work is thus complementary to the  previous WFC survey analysis reported~\citep{Verrecchia} and it is a natural evolution of the work developed for the IBIS/INTEGRAL survey catalogue~\citep{Bird4}. 
Indeed, the WFC principal characteristics are directly comparable with those of IBIS~\citep{Ubertini}, the coded mask gamma-ray telescope on board the {\it INTEGRAL} satellite~\citep{Winkler}. The aim of this paper is therefore  to apply the same techniques developed for the IBIS survey~\citep{Bird1,Bird2,Bird3,Bird4} to the BeppoSAX WFC data, searching for faint persistent sources in the total mosaic maps made from individual WFC sky images. IBIS and the SAX WFCs have a complementary and partially overlapping energy range (3-28 keV for WFCs and 17 keV - 1 MeV for IBIS), allowing studies of persistent sources over a larger energy range.  
 Results from this work have also been used to give an independent check of some of the fainter sources detected in the IBIS survey catalogue production: details of the correlation between IBIS and WFC detections can be found in~\citet{Cap2009}


 \begin{table}[!]
 \begin{center}
\caption{{\it BeppoSAX} WFCs principal characteristics.}
 \label{charact}
 \begin{tabular}{cc}
 \hline
 \hline
  parameter & WFC value \\
 \hline
 \hline
 Energy range & 3-28 keV \\
 \hline
 Energy resolution & 20\% at 6 keV \\
 \hline
 Effective area & 140 $cm^{2} $\\
 \hline
 Field of view & $40^{\circ}\times 40^{\circ}$ (FWZR)\\
                &          $20^{\circ}\times 20^{\circ}$ (FWHM)   \\
 \hline
 Angular Resolving Power & 5'\\
 \hline
 Source location accuracy & $<$ 1' \\
 \hline
 Sensitivity  in  $10^{5}s$ &  $\sim$1 mCrab (3-28 keV)\\
 \hline
 \end{tabular}
 
 \end{center}
 \end{table}

\section{WFC Sky Map Production}
\label{sec2} 
The WFC data are organised in short observational periods (OPs) of at least 100 ks. We collected all the available data from different archives and analysed all the collected OPs with the WFC Data Analysis System, extracting images in the 3-17 keV and 18-28 keV energy ranges.  The data analysis up to and including the image level has been performed with the final version of the WFC standard data analysis software\footnote{www.asdc.asi.it/bepposax/software/index.html}~\citep{Jager2} using the reference catalogue included in the software package.

 The WFC standard software uses the IROS method (Iterative Removal of Sources) to extract sources from the shadowgrams of the WFCs coded masks \citep{Zand}. During this procedure we used specific parameters included in the standard software in order to keep all the detected sources in the second and subsequent  IROS iterations indepenently of the source identification (`-m4' option within the standard software). 
 Moreover we lowered the IROS threshold (allowing up to 300 iterations), such that the source removal continued further into the low significance detections and/or noise than would normally be done for transient detection, retaining low significance information for the mosaicing process.

We subsequently applied a filter in order to eliminate corrupted and noisy images  by comparing the root main square (rms) and mean of each flux image against the average image rms and image mean level derived from all images. Any images with an rms or mean level more than 5 sigma away from the nominal average values were not incorporated into the final mosaic. After the filtering procedure,  about 95\% of the total initial number of OPs were used in the subsequent analysis.

An all-sky mosaic of the images has been generated using the same software used for the IBIS survey~\citep{Bird1,Bird2,Bird3,Bird4}. The mosaicing procedure is designed to average a large number of small single images that cover almost all the sky into a single all-sky image.
 Thus the images of the filtered OPs have been combined together into all-sky mosaics.   For each sky pixel, the mosaic software establishes a weighted mean flux, weighting each input image contribution according to the variance of the signal in each input pixel. The events are fully redistributed in the final sky map
pixels that oversample the original image pixels and system PSF. The process generates flux, error, significance and exposure mosaic images.

  This procedure is strongly tuned towards the detection of persistent (and weak) sources. Even if a source field has been observed for a long time, intrinsic variability of a source may mean that it is detected only in few individual images, and it will not be detectable in the final map.  

The signal to noise ratio of a persistent source will increase with the number of added images. Conversely, noise and imaging artefacts in individual OP images that would create false detections are, by their very nature, detected in the same positions in only one image, and will be lost into the mosaic background provided that two conditions are satisfied:
\begin{itemize}
\item many images must be summed together (we assume a conservative minimum value for the exposure of 1.4 Ms - at least 14 images considering that the maximum possible exposure of a single image is 100 ks). In the regions of the map with an exposure less than this value, we did not consider any detection that was not reported before in the catalogue of ~\citet{Verrecchia}. 

\item the averaging procedure could fail to eliminate the false detections if they are due to systematic image artefacts resulting from the image reconstruction process if the sky pointing direction of the telescope is repeated. In this way, systematic effects will appear in the same sky positions and will be summed during the mosaic process. However, this is not the case of the WFC images due to their random pointings deriving from the serendipitous nature of the observing programme (this kind of effect cannot be totally excluded for the galactic centre region, details in Sections~\ref{noise} and~\ref{saxj_wfc}) 
\end{itemize}


 Moreover the mosaic procedure, as it combines many images from many pointings, averages any differences due to the off-axis response that are not completely corrected by the software (see \citet{Verrecchia} for details).
Also the Earth occultation simply contributes to the background of the final mosaic.

\begin{figure}[!h]
\centering
\includegraphics[angle=0,scale=0.83]{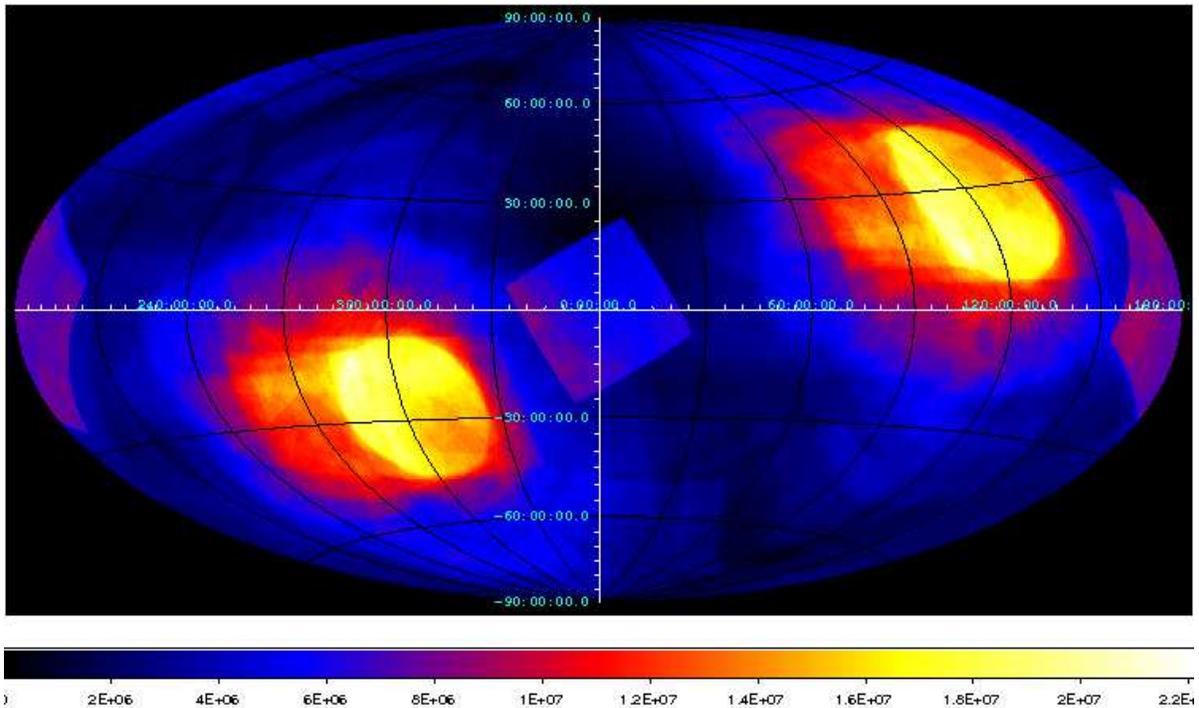}
\caption{\small{The WFC mosaic exposure map (in seconds) in Galactic Coordinates. The two zones with higher exposure are due to the polar passages of the Extended Science mode one (ESM1) and two (ESM2)~\citep{Tramutola}; the high exposures on the Galactic Centre and the  Anti-center zones  are due to the WFC core program observations of the Galactic Centre region. }}
\label{expmap}
\end{figure}

The collected data covered all the sky - although not uniformly. In fact, as Figure~\ref{expmap} shows, there are zones with much higher exposure such as the Galactic center and the fields centered on  the North and South Celestial poles due to the observing strategy and manouvering techniques of the {\it BeppoSAX} satellite~\citep{Tramutola}. The average exposure over the entire map is about $4\times 10^6$ s, even if there are some regions with an exposure of  two orders of magnitude less and others where the exposure reaches $\sim$ $2\times 10^7$s. There are two principal regions of low exposure: the zone around Sco X-1 (Sco X-1 was so bright that all the data containing  this source in the field of view were corrupted) and two small zones of about 5 degree radius 180 degrees apart at coordinates  (l,b)=(120.2$^{\circ}$, -57.4$^{\circ}$) and (299.0$^{\circ}$, 68.2$^{\circ}$).

The higher energy range map (18-28 keV) is affected by more noise and larger distortion of the Point Spread Function (see Section~\ref{noise}). Thus for source searching we only use the 3-17 keV map. 

\begin{figure}[h!]
\centering
\includegraphics[angle=0,scale=0.89]{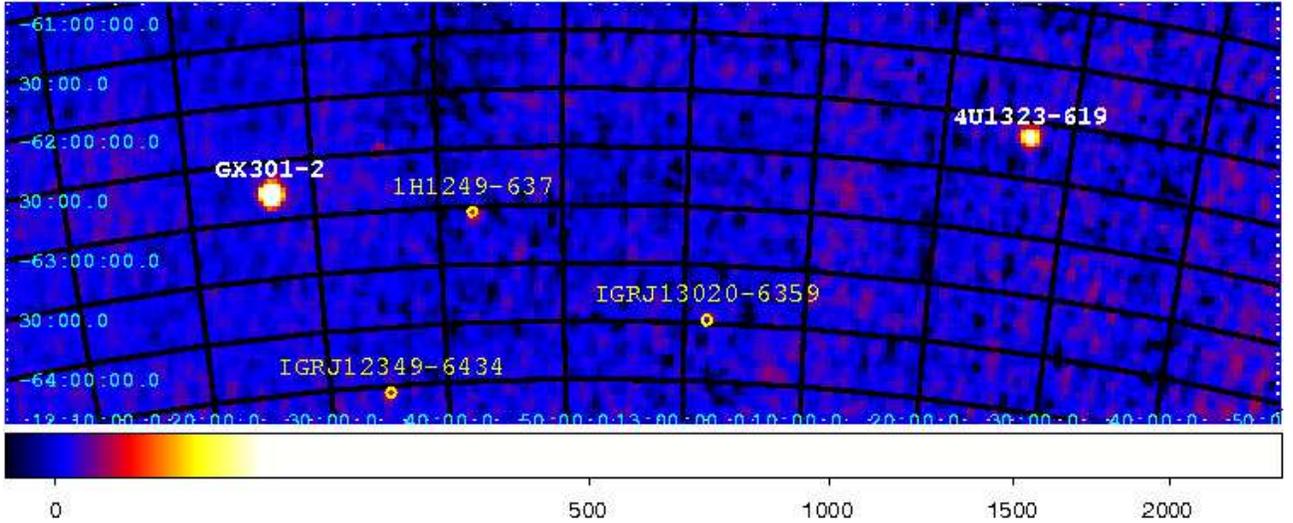}
\caption{Zoom of the WFC final mosaic map  (in units of sigma) centred around the GX301-2 field. The sources indicated in yellow have been reported only by our catalogue, while those with white labels have been reported in both our and \citet{Verrecchia} catalogues}
\label{Zoom}
\end{figure}

 It is clear that the analysis of WFC data carried out so far has  
concentrated on locating transient sources, whereas our methods allow a much more efficient detection of weak persistent sources:  as an example Figure~\ref{Zoom} shows a zoom of the all-sky WFC map
(3-17 keV) around GX301-2: the sources indicated with white labels have been reported in both our and \citet{Verrecchia} catalogues, while the sources labelled in yellow have been reported only in our catalogue. The three sources with yellow labels are classified as faint persistent (even if variable) sources (see e.g. \citet{Bird4}). The flux of the sources present in Figure~\ref{Zoom} in both catalogues are reported in Table~\ref{tab2.1}

\begin{table}[!h]
 \begin{center}
\caption{Fluxes of sources shown in the sky field of Figure~\ref{Zoom}. flux$_{ver}$ is the average flux extrapolated from~\citet{Verrecchia} (the large errors represent the source variability rather than an intrinsic measurement uncertainty), flux$_{mosa}$ is the flux derived from the all-sky mosaic map taken from Table~\ref{tab:sources}.}
 \label{tab2.1}
 \begin{tabular}{ccc}
 \hline
 \hline
source name & flux$_{mosa}$ & flux$_{ver}$\\
- &  (3-17 keV)mCrab & (2-10 keV) mCrab \\
GX301-2 &  20.3 $\pm$ 2.1& 18$\pm$26 \\
4U 1323-619 &  4.9 $\pm$ 0.5 & 7$\pm$5\\
IGR J12349-6434 & 0.6 $\pm$0.1 & -\\
IGR J13020-6359 & 0.4$\pm$0.1 & -\\
H 1249-637   &  0.7$\pm$ 0.1 & -\\


\hline
 
\end{tabular}
 
 \end{center}
 \end{table}

\begin{figure}[!h]
\centering
\includegraphics[angle=0,scale=0.89]{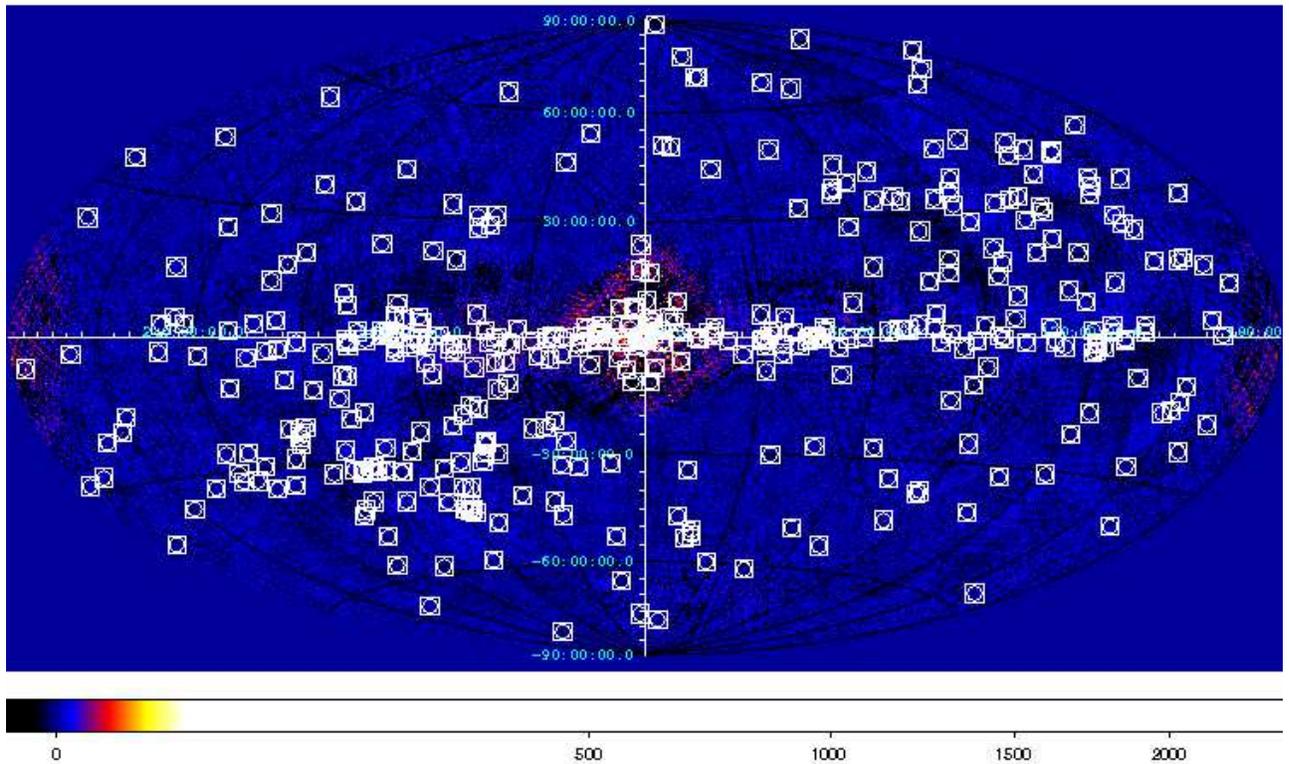}
\caption{WFCs final mosaic between 3-17  keV  in units of sigma. The symbols represent the detected sources positions. The parts of the map 
around the Crab and the Galactic Centre, 
present a higher level of noise.}
\label{skyCrosses}
\end{figure}

 Finally, Figure~\ref{skyCrosses} shows the 3-17 keV WFC final sky map  in units of $\sigma$. The  symbols represent the detected source positions, the parts of the map in red are zones with high level of noise (i.e.  the Galactic and Anti-Galactic Centre).

\subsection{Image problems - the `comet' effect and noisy regions}
\label{noise}


The IROS procedure performs a cross-correlation  between the detector image and the mask pattern via a matrix multiplication during every iteration and the detections are localised by fitting any peaks with an expected Point Spread Function (PSF). After the positions are compared with a reference catalogue,  the effect of sources on the detector plane are then simulated and subtracted.  Because each source is not simulated in exactly the same position in each single OP image~\citep{Zand}, this results in a slight broadening of the final PSF in the mosaic image.

The PSF can also be different from one single observation to the next as a result of photon penetration into the WFC detector gas chamber. 
This effect becomes more evident both at higher energy ranges and at large off-axis angle detections, and hence can change between OPs as the pointing direction changes.

 In fact as reported in \citet{Zand}, the photons can be absorbed at any depth {\it d} within the WFC detector. The probability of absorption of a photon in a $\Delta d$ thick layer at a depth $d$ is proportional to:\\
\begin{math}
P(d) \propto e^{-d/l(E)} T(d) \Delta d;
\end{math}\\
 where l(E) is the mean free path of a photon with an energy E and T(d) represents the blocking by the three WFC detector wire grids and the cut-off due to the finite detector thickness. \\
The projection of $P(d)$ on to the sky plane influences the PSF. This takes the form of $D(E)$tan$\alpha$ where $D(E)$ is the maximum depth for photons of energy E and $\alpha$ is the off-axis angle. \\
Thus the photon distribution projected on the detector plane at energies above $\sim$15 keV shows an asymmetric tail which cuts off at positions corresponding to that of the grid planes and to the bottom of the detector (i.e. 3 mm on the detector plane for a source at an off-axis angle of 4$^{\circ}$ at 30 keV. See~\citet{Zand} for further details).

\begin{figure}[!h]
\centering
\includegraphics[angle=0,scale=0.8]{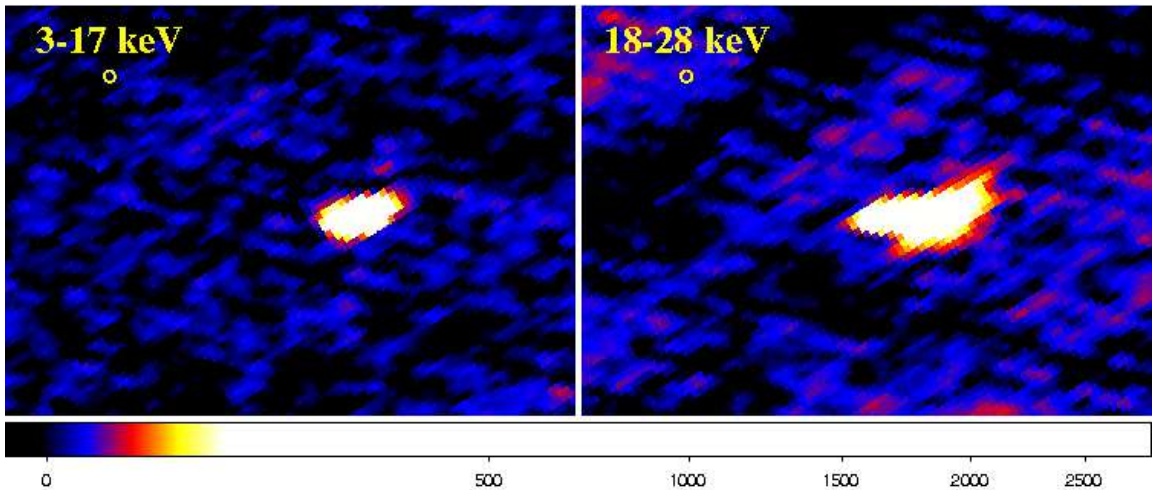}
\caption{\small{  Zoom of the final mosaic  map (expressed in sigma) on the Crab nebula  in the two energy ranges  (left: 3-17 keV, right: 18-28 keV) showing that the distortion of the PSF (`Comet effect') is more evident at high energies.}}
\label{comets}
\end{figure}

These effects have a significant impact on the source PSFs in the total map, creating a sort of `comet effect' in the PSF shape that becomes worse at higher energies, as Figure~\ref{comets} shows. Thus for source searching we choose the 3-17 keV energy range in order to both maximise the instrument sensitivity and minimize the deformation of the PSF in the maps.

Following the failure of the gyroscopes on board the satellite, the star trackers were used to control the rotations and the pointing of the satellite~\citep{Tramutola}. This procedure required those zones of the sky regularly observed (such as the Galactic Centre region and consequently the anti-galactic centre and the zone around the Crab nebula) to be observed with the satellite (and WFC detector) in the same position with respect to the sky, thus with the same bright stars in the star tracker field of view.  As the {\it INTEGRAL} pointing strategy has demonstrated for coded mask instruments, observing the same sky zone with different pointing configurations can significantly reduce the background~\citep{Courvoisier} image artefacts.
For this reason, the Galactic Centre zone (and the anti-Centre) have a significantly higher background in the WFC  total mosaic map compared to the other sky zones where the serendipitous nature of the pointings have the same effect as the {\it INTEGRAL} pointing strategy. Moreover the Galactic centre zone has an intrinsically higher noise also in a single image, due to the presence of a large number of sources in the field of view which make the image deconvolution more difficult.

\section{The Table}

\subsection{Source List Generation}

Two methods were used to create an initial source list. The first used a tool developed for the IBIS survey~\citep{Bird3}  that searched for excesses exceeding a local threshold set by a baseline statistical threshold scaled by the local rms fluctuations within the map. This tool is intended to suppress the detection of fake sources in areas of the map with high non-statistical fluctuations. A second method based on SExtractor 2.4.4 software~\citep{Sex} has been used to cross check the results with a bandpass filter (Gauss filter) to minimize the source confusion in crowded fields.  The list of excesses was then checked manually. A baseline acceptance limit at 4.8 sigma level has been adopted, although we stress again that the effective threshold in areas of high systematic artefacts will be considerably higher thus the acceptance limit varies considerably from zone to zone of the map. For example in the Galactic Centre region the local acceptance limit is at  $\sim$12$\sigma$. After acceptance, the source positions and fluxes were evaluated using a barycentering method to determine the centroid of the source profile. The mean flux of the sources was determined from the count rate at the position of the source maximum significance ( the counts-to-flux conversion is obtained assuming a Crab-like spectrum for the sources). 
After all the checks, the final list contains 458 excesses above 4.8 sigma. Of these excesses, 404 were  identified  as sources while 54 were considered to be map artefacts because of the PSF shape or the proximity to an image structure. The sources in the final list were then classified by a process of correlation with other existing catalogues.

\subsection{Position error}

\begin{figure}[!h]
\centering
\includegraphics[angle=0,scale=0.6]{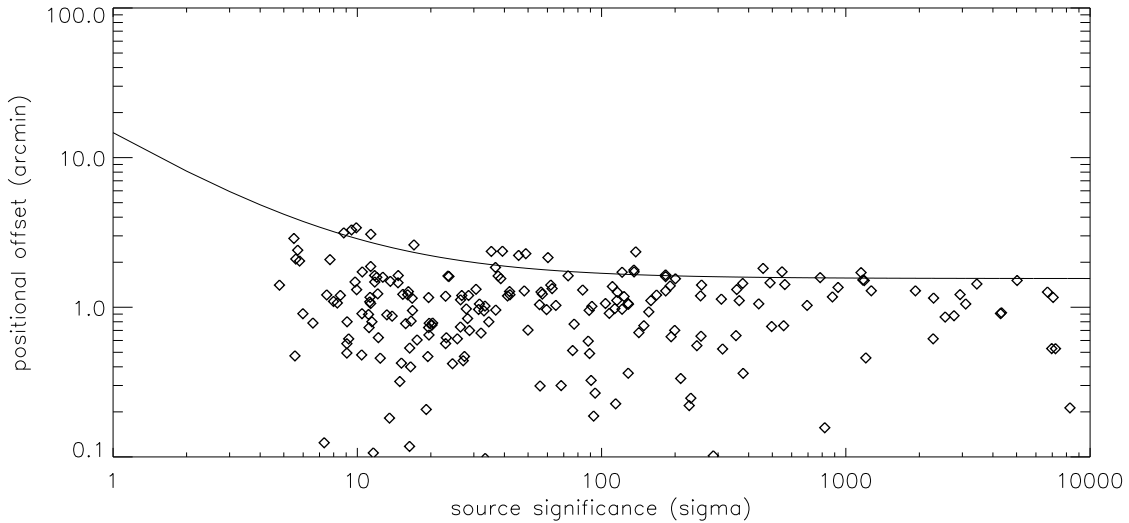}
\caption{ Offsets between the WFC map positions and IGRC positions as a function of source SNR. The curve represents the extrapolated error (90\% confidence limit).}
\label{errore}
\end{figure}

The PSF distortion due to the `comet effect' and the presence of other systematic effects prevent us from simply extrapolating the point location accuracy from the mosaic maps with a fixed confidence level.
In order to estimate the source location error  radius of the WFC sources, we compared the positions of the known sources detected in the map with their best known positions, applying a procedure similar to the one reported in \citet{Scaringi}. 
The best positions were extracted from the {\it INTEGRAL general reference catalogue (IGRC)}, considered to be one of the most recent and comprehensive compilations of accurate X-ray and hard X-ray positions~\citep{ebisawa}.  We have taken into account only those sources in the IGRC with a position specified to better than 30\arcsec, i.e. those providing a well-defined 'reference' position; this was possible for a total of 204 sources detected by the WFC.   
Following the procedure described in~\citet{Scaringi}, we plotted the offset between our positions and the reference ones, as a function of the WFCs Signal to Noise Ratio (SNR) for these 204 sources, after which we binned the offset values in order to have enough statistical significance for each bin.
 This method can take into account both the systematic and the statistical errors as it makes no assumptions about the form of the position error.
The best fit curve for the 90\% confidence radius, plotted in Figure~\ref{errore}, is: %
\begin{math}
Y={A(0)\over X} e^{A(1)} + {A(2)}; 
\end{math}
where: 
A(0)=3.21; A(1)=1.41; A(2)= 1.56;
The errors reported in Table~\ref{tab:sources} are extrapolated from Figure~\ref{errore}.

 It is important to note that the best positions  for variable and transient sources are generally derived from the single observation with the highest sigma. The positions derived from the overall mosaic for these sources are essentially degraded by the many non-detections added when forming the final mosaic. Conversely, for faint persistent sources, the position extracted from the mosaics is the best that can be obtained, and has the lowest error.

\subsection{Comparison with the previous WFC catalogue}
\label{lcgen}

A WFC catalogue of sources was published by Verrecchia et al. in July 2007. That work is based on the analysis of each single pointing and is optimised for transient source detection (a similar work, restricted to the Galactic plane zone, was published in \citet{cap2004}). Our work is instead based on the searching of mosaic maps and is primarily intended to identify persistent sources not necessarily visible in a single OP. For this reason our list of sources is somewhat different from the one published in \citet{Verrecchia}.

\citet{Verrecchia} lists 253 sources while our catalogue contains 404 sources; 222 sources are present in both catalogues. The 31 sources reported only in \citet{Verrecchia} are all faint sources detected in only one or two individual OPs, thus as we expected they are not detected in the total map. In particular, within the 11 brightest sources listed only by \citet{Verrecchia}, there are 5 sources for which there is a detection in the total map but the detection level is lower than 4.8 sigma. The other 6 sources are located near structures in our map or in noisy regions (like the Galactic centre region) where the detection threshold is high.

As we expected, the 182 sources found only by us (shown in bold in Table~\ref{tab:sources}) are mostly persistent or   quasi persistent. It is important to note that within these 182 sources, there are 17 IGR sources and 9 {\it Swift} sources that have been discovered only after the end of the Beppo SAX mission.  Moreover, the list contains ten new sources.

In order to verify and further quantify the intrinsic variability of the source populations in the two catalogues, we have inspected the statistical properties of light curves extract for each source. We developed  a tool for WFC light curve production, that reads the flux (counts/s/cm$^{2}$) from the source position in each single WFC pointing image using the {\it The IDL Astronomy User's Library procedures}\footnote{http://idlastro.gsfc.nasa.gov/}. Unfortunately, any  tool such as this will be affected by the problem reported by~\citet{Verrecchia}: the WFCs standard software does not totally correct the differences in flux due to the different position of the sources in the field of view of each WFC pointing. 
This effect adds a systematic error of about 10\%. 

\begin{figure}[!h]
\centering
\includegraphics[angle=0,scale=0.5]{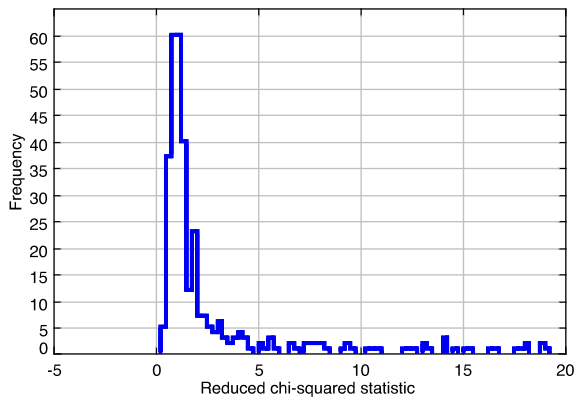}
\caption{Variability distribution of the catalogue sources, shown in the form of the reduced $\chi^{2}$ when the light curve is compared to a model of a constant mean flux.}
\label{chi}
\end{figure}

\begin{figure}[!h]
\centering
\includegraphics[angle=0,scale=0.5]{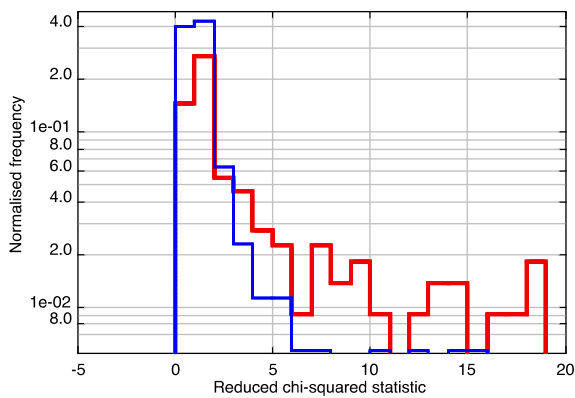}
\caption{ Blue histogram: distribution of the reduced  $\chi^{2}$ of the sources detected only in our catalogue. Red histogram: distribution of the reduced  $\chi^{2}$ of the sources detected in both our and the previous catalogue}
\label{vari_per}
\end{figure}

 The intrinsic variability for each source in our dataset was determined by performing a check for excess variance (i.e a chi-squared test against a constant mean flux) for each of the light curves. As expected, the $\chi^{2}$ distribution is peaked around 1 (indicating a persistent source) with a long tail representing the variable sources (see Figure~\ref{chi}). Moreover, as Figure~\ref{vari_per} shows, comparing the variability statistic of both the sources detected also in the previous catalogue and the sources detected only in this catalogue, it is clear that the former are intrinsically more variable. 

\subsection{Flux errors}

As discussed in the previous section, the image reconstruction algorithm does not make a full correction for the off-axis response of the WFC cameras, resulting in a sytematic increase in the flux uncertainty of $\sim10$\% in any given OP.

\begin{figure}[!h]
\centering
\includegraphics[angle=0,scale=0.45]{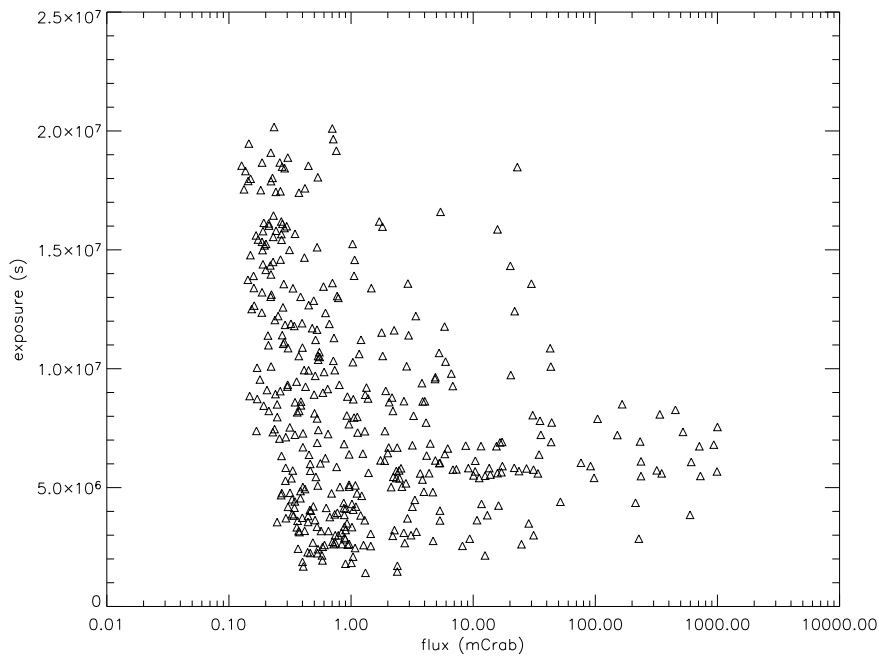}
\caption{  Exposure versus flux for each detected source. The sensitivity limit of the survey is clearly visible as the left border of the detected sources.}
\label{flux_vs_expo}
\end{figure}

The mosaicing process generally reduces the effects of this poorly corrected off-axis response by averaging the fluxes from many OPs taken with different pointings, meaning that the source flux is measured at many different off-axis angles. However, in some cases where, for example, the single observations are not pointed with a random configuration (as in the galactic centre region) or when the source is highly variable, the fluxes in the maps could be affected by  a systematic error higher than the estimated 10\% . However, a plot of the source fluxes versus exposures, as in Figure~\ref{flux_vs_expo}, does not indicate any source flux anomalies. Thus the flux errors reported in Table~\ref{tab:sources} can be considered to include both the statistical and the systematic errors.

%
%
%

\section{The source sample}

\begin{figure}[!h]
\centering
\includegraphics[angle=0,scale=0.35]{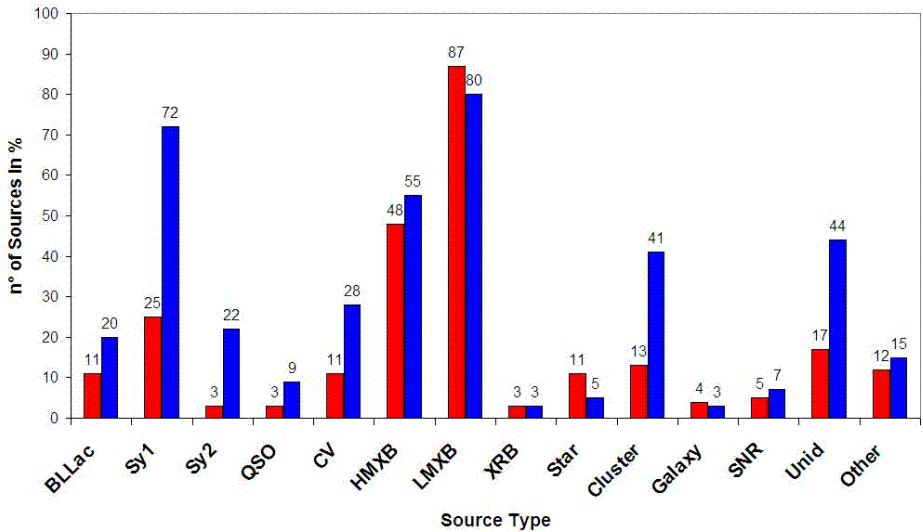}
\caption{Source type distribution for this work (blue columns) compared with the one reported in  \citet{Verrecchia} (red columns). }
\label{isto}
\end{figure}

 In terms of the sources themselves, the WFC source type distribution from this work is significantly different from the one reported by the previous WFC catalogue and reflects the higher fraction of persistent sources in our catalog.
This indicates that one must always be aware of the timescale of the sensitivity of surveys when using them - the hard X-ray sky varies on many timescales - and performing a source search on any one timescale inevitably introduces a bias towards different source types.
In fact, as Figure~\ref{isto} shows, our catalogue has a higher number of typical persistent objects like Seyfert 1 (Sy1), Seyfert 2 (Sy2) and Clusters of galaxies. In particular,  compared to the previous catalogue, the number of Seyfert galaxies is increased by about 3 times for Sy1 and about 7 times for Sy2. While the number of Clusters of Galaxies detected increases by about 3 times.

 It is also noticeable that our catalog contains a higher number of unidentified sources with a distribution  that traces the higher exposure zones of the map.

The low mass X-ray binaries are the most populated class of sources of this catalogue. Reflecting the highly variable nature of these objects, their number is slightly lower than the one previously reported (the total mosaic map lost the faintest transient objects) and all of them have been also detected in ~\citet{Verrecchia} except for  1E 1743.1-2843 that is a faint (0.1-0.2 L$_{E}$) and peculiar source classified as a LMXB  with a persistent nature \citep{DelSanto}.   1E1743.1-2843 is a typical source for which our work is optimised. 

The High Mass X-ray Binaries are mostly the same as those observed in the previous catalogue. The new detections (7 sources) are all Be/X-ray binary systems in which we caught either persistent or long-term outburst emission. 

We detected 17 new Cataclismic Variables; looking at the different types of CV, mostly all the new objects found are IP (14 object for a total of 17), the CV subtype with the harder spectra ( see e.g. ~\citet{Scaringi2}), while there are 3 new detections of Dwarf Novae out of a total of 5 sources (for a discussion on the hard X-ray spectra of DNs see Landi et al. in preparation).

\begin{figure}[!h]
\centering
\includegraphics[angle=0,scale=0.35]{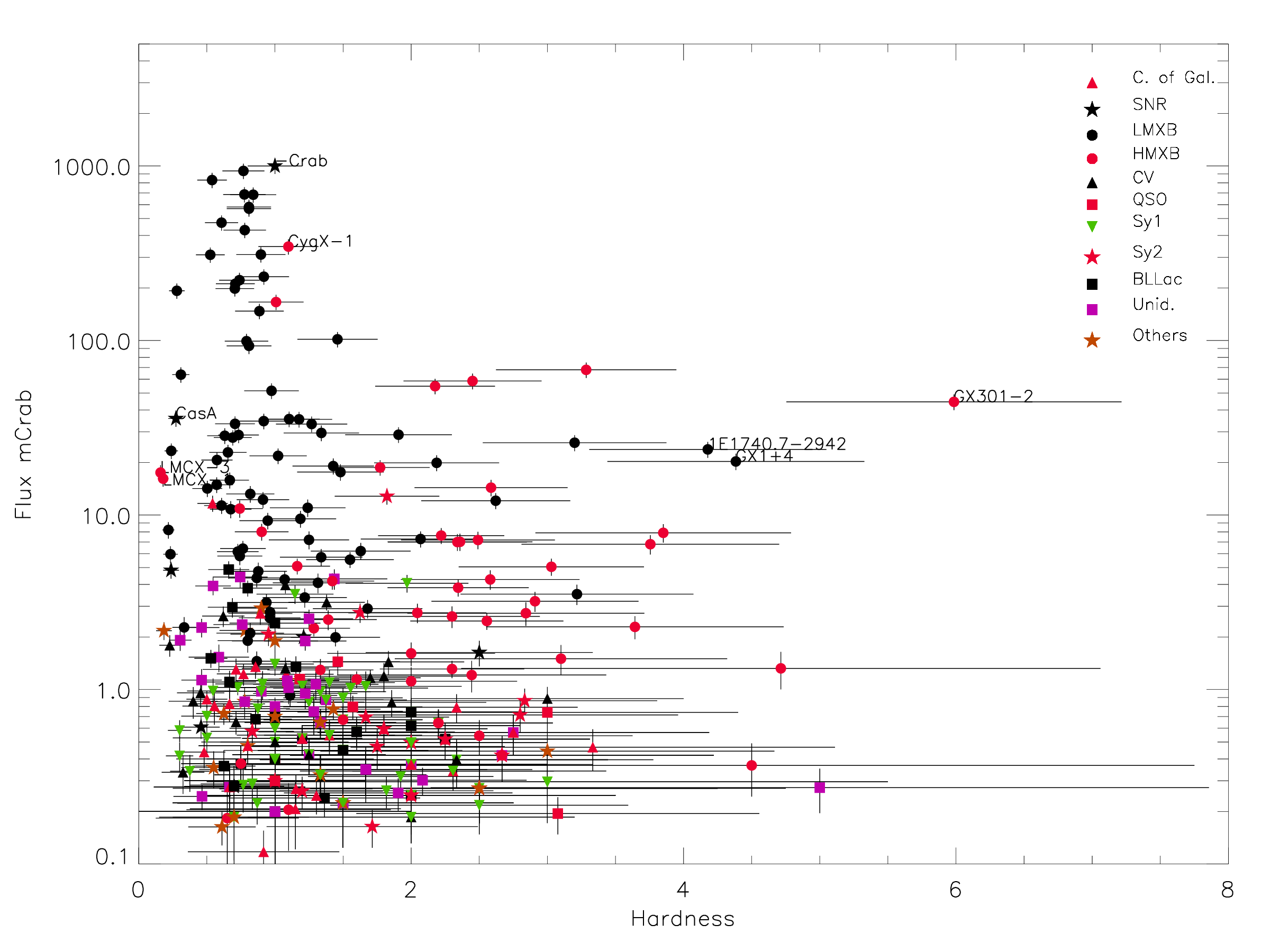}
\caption{hardness ratio vs flux in mCrab. The hardness ratio is defined as Flux$_{17-28keV}$/Flux$_{3-17keV}$}
\label{hardness}
\end{figure}

Figure~\ref{hardness} shows the averaged hardness ratio (HR) of all the sources, extracted from both 3-17 keV and 17-28 keV mosaic maps. Even if the large errors only allow the extraction of basic information for faint sources, we can speculate about the global behaviour of the luminous sources.
 
The bright source sample is mostly formed by XRB; as Figure~\ref{hardness} shows, the HMXB are the hardest emitters: nine of them show a HR value greater than 3. The two softest HMXB are indeed LMC X-1 and LMC X-3, both of them are persistent sources often detected with a bright disc black body component~\citep{Yao}.
The hardest source of the entire sample is the HMXB GX301-2, one of the most massive X-ray binary known (see e.g ~\citet{Leahy}), while the hardest LMXB is 1E1740.7-2942 (a persistent source often in hard state, ~\citet{Bouchet}) and GX1+4 an accreting X-ray pulsar~\citep{Ferrigno}. 
However, for the most part, the LMXB have HR values that lie between 0.5 and 1.5. In particular the most luminous LMXBs show HR values below 1. This is in agreement with the expected behaviour of a LMXB which generally is very bright during its soft state~\citep{McC}. 
The huge differences between the XRB hardness ratios are not only due to the intrinsic properties of these sources but are also due to the specific observations. With the XRBs being extremely variable, the averaged HR values depend strongly on the source spectral states at the time of the WFC observations and may not represent a real time-average hardness value.

 We produced the logN-logS for the most populous source types present in our catalogue: the LMXB and HMXB. The distribution is only indicative because even if the sky coverage is virtually complete, the exposure is not uniform. 

\begin{figure}[!h]
\centering
\includegraphics[angle=0,scale=0.55]{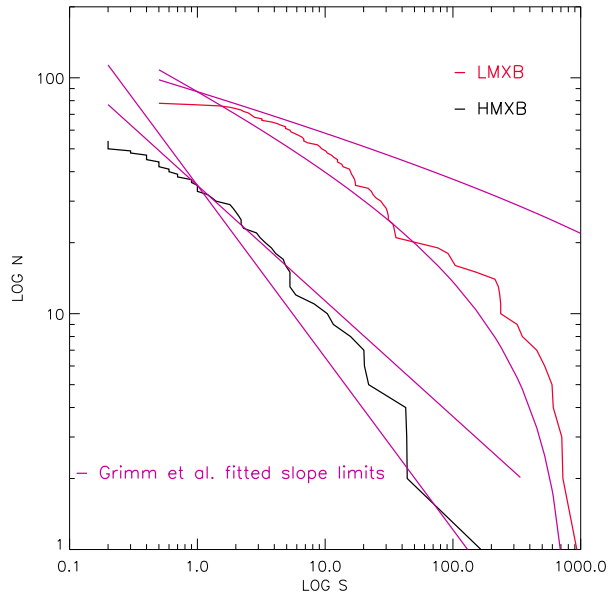}
\caption{ logN-logS distribution of respectively HMXB (black line) and LMXB (red line). The straight lines represent the lower and the upper limits of the best fit slopes reported by \citet{Grimm}. The fluxes are expressed in mCrab. }
\label{lognlogsBin}
\end{figure}

Our logN-logS distribution is consistent with the one reported in ~\citet{Grimm} both for HMXB and for LMXB. As Figure~\ref{lognlogsBin} shows, the latter present a cutoff and a flatter behaviour with respect to the former. The  straight lines in Figure~\ref{lognlogsBin} represent the lower and the upper limits of the best fit slopes reported by ~\citet{Grimm}.  Our plot, above about two mCrab, is in good agreement with ~\citet{Grimm} even if, in spite of a total sky coverage, no exposure correction has been added to our data: the exposure spans from 2$\times$10$^{6}$s to 20$\times$10$^{6}$s for LMXBs and 2.5$\times$10$^{6}$s to 15$\times$10$^{6}$s for HMXBs respectively. 

\subsection{Searching for transient sources with light curves}
\label{curve}

\begin{figure}[!h]
\centering
\includegraphics[angle=90,scale=0.4]{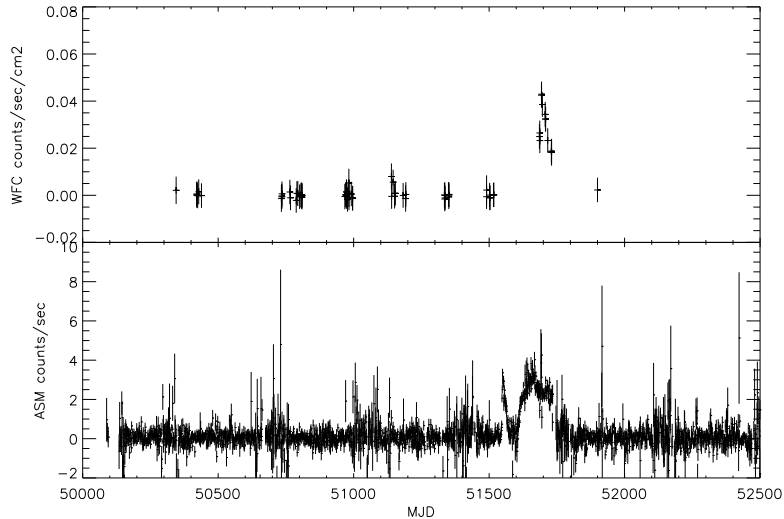}
\caption{XTE J1118-480: top panel: WFC 3-17 keV light curve. Bottom panel: RXTE/ASM monitor light curve (1.3-12 keV)}
\label{minni}
\end{figure}

 As discussed in section~\ref{lcgen}, we are able to extract (with some limitations) light curves for any point on the sky by extracting fluxes from the OP images. This method can be used to perform an additional search for and analysis of known sources, since it only needs the position of the selected source. As an example Figure~\ref{minni} shows both WFC and ASM/RXTE light curves of the LMXB system XTE J1118+480.

This light curve tool is also useful to search for transient sources that are below the detection threshold in the total mosaic but are also too faint to be clearly detected in single OP images, falling somewhere between the capabilities of the two catalogues search methods. A good example of this is the case of IGR J17091-3624~\citep{Zand1}. This transient X-ray binary is not detected in the total mosaic but mosaicing only the WFCs observations near the known outburst periods (September 1996 and September 2001) the source is clearly detected (Capitanio, PhD Thesis 2007). Obviously this procedure implies that the position of the source and its outburst date(s) must be previously known, and this unfortunately significantly limits the application of this method. 


\subsection{The new source candidates}


The new source candidates detected in the maps were chosen using three principal conservative criteria: a signal to noise ratio greater than 12 ( set to cope with even the worst local sigma level in the mosaic map), an exposure higher than  4$\times$10$^{6}$s, the average exposure over the entire map, (see Section~\ref{sec2}) and a light curve that does not present any spikes in single images.  In fact we visually inspected several possible new transients and rejected all of them on the basis of their being at the image border or near a structure or presenting an unacceptable PSF. With our work being specfically optimised for faint persistent sources, we did not expect any new transient sources with respect to the previous catalogue that was optimised for transient detection. Applying these criteria, 10 new source candidates have been selected (3\% of the total number of sources).

We searched for these 10 new sources within the X-ray observation archives and through catalogues from other energy ranges. The fields containing the new WFC sources have never been observed in the X-ray energy range below 20 keV  (except for the ROSAT all sky survey that did not detect them) and they have not been detected above 20 keV in either Swift/BAT or IBIS/ISGRI mosaics.

 Confidence in these 10 new detections is strongly supported by the many other sources uncovered by the same technique that proved to be correlated with known sources. Of course in this kind of work, it is always possible that some false detections will be included, and the new sources can only be truly verified by follow-up observations or detections in other instruments.

 Table~\ref{tab:newsources} summarises the principal characteristics of these 10 sources; in particular the last column reports the possible radio and infra-red counterparts of some of the sources.

Another source, WFC J1818-1658, was initially added to the list of the new sources, but after a more accurate analysis it was identified as the Supergiant fast X-ray transient SAX J1818-1703, as described below. 

\subsubsection{The curious case of WFC J1818-1658/SAX J1818.6-1703}
\label{saxj_wfc}

SAX J1818.6-1703 is an anomaly within the WFC catalogs.  This source, discovered by the WFC in 1998~\citep{Zand_sax}, was not automatically recognised in our mosaic map, and it is not reported in~\citet{Verrecchia}.

The best position found in the WFC map for our candidate WFC J1818-1658 lies 9.8$\arcmin$ away from SAX J1818.6-1703  with a calculated error radius of 2$\arcmin$, the source being detected at the 32.5 sigma level (in a sky zone with a local averaged background of about 10 sigma level). The source was automatically classified as a new source. 
A counterpart search in all  {\it Swift}/XRT and XMM-Newton/EPIC data with the source in the field of view, did not discover any plausible counterpart  within the 2$\arcmin$ radius error circle. Although the position of WFC J1818-1658 is consistent with a serendipitous {\it XMM} source, 2XMMi J181813.9-165724,  detected by both {\it Swift}/XRT and XMM-Newton/EPIC, the flux expected for a 32.5 sigma level source in the WFC mosaic map should be 100 times greater than that of this faint XMM object.



 On the other hand, we know from IBIS studies~\citep{Bird_sax} that SAX J1818.6-1703  (with counterparts clearly detected in most of the XRT and XMM images analysed) appears in the IBIS persistent search mosaics as a result of occasional outbursts and low-level emission that occur during its periastron passages every 30d.

\begin{figure}[h!]
 \centering
\includegraphics[angle=0,scale=0.85]{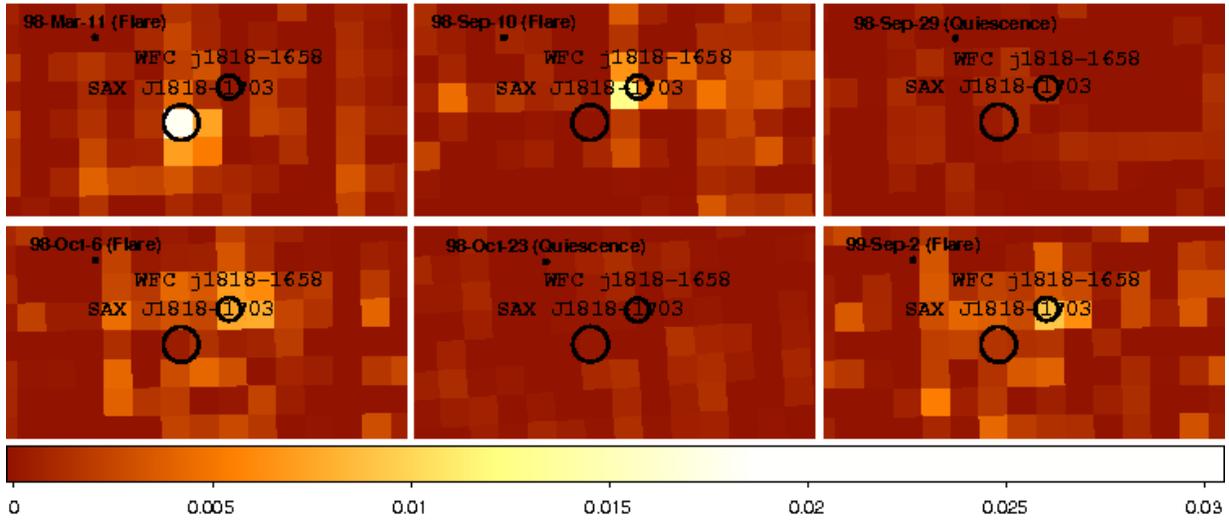}
 \caption{Selected WFC single-OP flux images,  expressed in $counts/s/cm^{2}$, of the SAX J1818-1703/WFC J1818-1658 region during both periodic flaring activity and quiescent states. }
 \label{saxj}
 \end{figure}

Looking at each single WFC observation after the first detection of the source~\citep{Zand_sax}, there is a faint transient object that appears recurrently in the data. This object has a position that slightly oscillates between the positions of the two sources  (SAX J1818.6-1703 and WFC J1818-1658). However  the periods in which the source is visible is recurrent and consistent with the known flaring period of SAX J1818-1703 (30$\pm$ 1 days)~\citep{Bird_sax,ZurH}. Figure~\ref{saxj} shows some WFC single OP detections during both the flaring and the quiescent periods. 

We can say that the timing analysis indicates that WFC J1818-1658 and SAX J1818-1703 have an high probability to be the same source and the shifted position of the SAX J1818.6-1703 is caused by the particular sky zone in which the source is situated. In fact this turns out to be a pathological case in which all the defects of the WFC map reported in the previous sections (i.e. Section~\ref{noise})  are manifest together: SAX J1818.6-1703 lies in the field of view of the periodic observations of the Galactic Centre region, thus the same star tracker configurations had been used each time. Moreover SAX J1818.6-1703 always lies at the border of these observations, thus at a huge off-axis angle.  We have used mean offsets throughout this work when discussing position errors, because we are aware that there can be a small subset of sources (like SAX J1818-1703) that have substantially larger errors. Quoting a 90\% confidence limit based on this subset would give a misleadingly high location error for the vast majority of our sources. Finally, except for the first detection ($\sim$100 mCrab) \citep{Zand_sax}, the source flares were very faint  ($\sim$ 1 mCrab), partly explaining why it has not been seen as a transient object, but is instead seen as a quasi-persistent emitter, just as it is in the IBIS survey analysis.

\section{Concluding remarks}


The wide field cameras on the BeppoSAX satellite were designed primarily to detect bright transient sources flaring within their large field of view. Despite this, the quality and quantity of data recorded has provided a legacy archive of quasi all-sky observations of the hard X-ray sky that has not been fully exploited.

We have successfully applied techniques developed for the {\it INTEGRAL}/IBIS survey to the BeppoSAX WFC dataset, on the basis that the two instruments are intrinsically similar in imaging method and operation. Our main aim has been to improve the sensitivity to weaker, more persistent sources not detected within individual WFC observations. The production and searching of mosaic maps from the ensemble of individual pointing is a good method to achieve this goal for persistent or quasi-persistent sources.

The success of this approach is evident in the detection of 182 sources not previously recorded in WFC catalogs. Most of these are known sources, partly because of the surge in hard X-ray detections in the {\it INTEGRAL}/{\it Swift} era, but around 35 of these sources would have been new discoveries for BeppoSAX if found at the end of the mission. Even though this work is partly limited by the optimisation of the WFC hardware and software for transient source searching, this represents a success for the approach. From a technical viewpoint, the areas that could still be improved include the flux reconstruction for individual pointings (and hence light curve production) and the PSF distortion that limits the source location accuracy and useful energy range.


When used in combination with more recent all-sky hard X-ray surveys, this catalog provides a view of this highly variable sky in another epoch with similar sensitivity (better than 1mCrab), and as such should be of value in any studies of variability in galactic and extra-galactic hard X-ray sources.

\clearpage
\begin{landscape}
\scriptsize
\renewcommand{\thefootnote}{\alph{footnote}}
\begin{center}
\begin{longtable}{cccccccccc}
\caption{BeppoSAX WFCs list of sources}
 \label{tab:sources} \\
\multicolumn{1}{c}{\textbf{Name\tablenotemark{a}}} &
\multicolumn{1}{c}{\textbf{Ra $(^{o})$}} &
\multicolumn{1}{c}{\textbf{dec $(^{o})$}} &
\multicolumn{1}{c}{\textbf{$\sigma$}} &
\multicolumn{1}{c}{\textbf{error (')\tablenotemark{b}}} &
\multicolumn{1}{c}{\textbf{flux$_{(3-17 keV)}$ \tablenotemark{c}}} &
\multicolumn{1}{c}{\textbf{flux$_{(17-28 keV)}$ \tablenotemark{c}}} &
\multicolumn{1}{c}{\textbf{source type\tablenotemark{d}}} &
\multicolumn{1}{c}{\textbf{sub type\tablenotemark{e}}} \\[1.0ex] \hline
\endfirsthead
\multicolumn{3}{c}{{\tablename} \thetable{} -- Continued} \\[1.0ex]
\multicolumn{1}{c}{\textbf{Name\tablenotemark{a}}} &
   \multicolumn{1}{c}{\textbf{Ra $(^{o})$}} &
    \multicolumn{1}{c}{\textbf{dec $(^{o})$}} &
   \multicolumn{1}{c}{\textbf{$\sigma$}} &
   \multicolumn{1}{c}{\textbf{error (')\tablenotemark{b}}} &
\multicolumn{1}{c}{\textbf{flux$_{(3-17 keV)}$\tablenotemark{c}}} &   
\multicolumn{1}{c}{\textbf{flux$_{(17-28 keV)}$ \tablenotemark{c}}} &
\multicolumn{1}{c}{\textbf{source type\tablenotemark{d}}} &
   \multicolumn{1}{c}{\textbf{sub type\tablenotemark{e}}} \\[0.5ex] \hline
\endhead
\multicolumn{3}{l}{{Continued on Next Page\ldots}} \\
\endfoot
 \\[-1.8ex] 
\endlastfoot
        {\bf IGR J00040+7020} &  0.913& 70.307&     7.7&   3.3& 0.2$\pm$ 0.1&     0.5$\pm$    0.1& Sy2 & - \\
                {\bf Mrk 335} &  1.591& 20.199&     7.4&   3.3& 0.4$\pm$ 0.1&     0.6$\pm$    0.1& Sy1 & - \\
  {\bf 1RXS J000635.7-690030} &  1.660&-68.979&     6.6&   3.5& 0.2$\pm$ 0.1&     0.4$\pm$    0.1& unid & - \\
          {\bf QSO B0014+810} &  4.298& 81.583&     5.9&   3.8& 0.13$\pm$ 0.03&     0.4$\pm$    0.1& QSO & -\\
                   4U 0022+63 &  6.333& 64.148&   189.6&   1.6& 5.9$\pm$ 0.6&     1.4$\pm$    0.2& SNR & - \\
                     V709 Cas &  7.177& 59.278&    31.6&   2.0& 1.0$\pm$ 0.1&     1.7$\pm$    0.2& CV & IP \\
        {\bf IGR J00335+6126} &  8.346& 61.428&     5.6&   3.9& 0.2$\pm$ 0.1&     0.2$\pm$    0.1& unid & - \\
  {\bf 1RXS J003422.2-790525} &  8.540&-79.077&     9.3&   3.0& 0.2$\pm$ 0.1&     0.5$\pm$    0.1& Sy1 & - \\
                1E S0033+59.5 &  8.920& 59.804&    39.3&   1.9& 1.3$\pm$ 0.2&     1.5$\pm$    0.2& BLLac & - \\
              IGR J00370+6122 &  9.256& 61.338&    11.8&   2.7& 0.4$\pm$ 0.1&     0.3$\pm$    0.1& HMXB & - \\
    {\bf RX J004241.6+411440} & 10.676& 41.242&     5.5&   3.9& 0.3$\pm$ 0.1&     0.2$\pm$    0.1& unid & - \\
                {\bf NGC 262} & 12.147& 31.951&     5.8&   3.8& 0.3$\pm$ 0.1&     0.8$\pm$    0.2& Sy2 & - \\
        {\bf RX J0051.3-7216} & 12.802&-72.247&     6.4&   3.6& 0.2$\pm$ 0.1&     0.13$\pm$    0.04& HMXB & - \\
                RX J0052-7319 & 13.084&-73.321&     8.7&   3.1& 0.3$\pm$ 0.1&     0.6$\pm$    0.1& HMXB & - \\
                       CF Tuc & 13.290&-74.647&     6.5&   3.6& 0.2$\pm$ 0.1&        -           & RSCVn & - \\
        {\bf XSS J00564+4548} & 13.838& 46.202&     6.2&   3.7& 0.3$\pm$ 0.1&     0.5$\pm$    0.1& CV & IP \\
                   gamma Cas. & 14.165& 60.702&   156.0&   1.6& 4.9$\pm$ 0.5&     5.7$\pm$    0.6& HMXB & Be \\
          {\bf XTE J0103-728} & 15.718&-72.740&     7.2&   3.4& 0.2$\pm$ 0.1&     0.3$\pm$    0.1& HMXB & Be \\
               PSR J0111-7317 & 17.850&-73.288&    57.2&   1.8& 1.8$\pm$ 0.2&     4.6$\pm$    0.5& HMXB & Be \\
              RX J0117.0-7326 & 19.292&-73.432&   496.7&   1.6& 15.8$\pm$ 1.6&    28.0$\pm$    2.9& HMXB & SG \\
                  3A 0114+650 & 19.510& 65.288&    92.8&   1.7& 2.9$\pm$ 0.3&     6.8$\pm$    0.8& HMXB & SG \\
                  4U 0115+634 & 19.667& 63.736&    33.0&   2.0& 1.0$\pm$ 0.1&     2.3$\pm$    0.3& HMXB & Be \\
     {\bf SWIFT J0123.9-5846} & 20.934&-58.820&    11.9&   2.7& 0.5$\pm$ 0.1&     0.5$\pm$    0.1& Sy1 & - \\
                      NGC 526 & 20.976&-35.067&    16.4&   2.4& 0.8$\pm$ 0.1&     1.0$\pm$    0.2& Sy1 & - \\
                  4U 0142+614 & 26.548& 61.747&    83.7&   1.7& 2.7$\pm$ 0.3&     0.5$\pm$    0.1& XRB & - \\
              RX J0146.9+6121 & 26.789& 61.342&    62.0&   1.8& 2.1$\pm$ 0.2&     2.7$\pm$    0.3& HMXB & Be \\
  {\bf 1RXS J015634.6-835836} & 29.167&-83.987&    12.4&   2.6& 0.3$\pm$ 0.1&     0.5$\pm$    0.1& unid & - \\
                 {\bf ICA 12} & 31.769&-74.429&     7.9&   3.2& 0.2$\pm$ 0.1&     0.4$\pm$    0.1& StC. & - \\
              IGR J02097+5222 & 32.418& 52.450&    10.4&   2.8& 0.4$\pm$ 0.1&     0.6$\pm$    0.1& Sy1 & - \\
            {\bf 1H 0215-007} & 33.634& -0.772&     7.7&   3.3& 0.4$\pm$ 0.1&     0.6$\pm$    0.1& Sy1 & - \\
           {\bf 1ES 0212+735} & 34.357& 73.801&     5.8&   3.8& 0.22$\pm$ 0.04&     0.3$\pm$    0.1& BLLac & - \\
               {\bf Mrk 1040} & 37.059& 31.308&     6.3&   3.6& 0.4$\pm$ 0.1&     0.8$\pm$    0.2& Sy1 & - \\
     {\bf SWIFT J0238.2-5213} & 39.595&-52.177&     8.3&   3.1& 0.3$\pm$ 0.1&     0.2$\pm$    0.1& Sy1 & - \\
  {\bf 1RXS J023832.6-311658} & 39.632&-31.267&     9.1&   3.0& 0.4$\pm$ 0.1&     0.22$\pm$    0.04& QSO? & - \\
         {\bf 1E 0236.6+6100} & 40.178& 61.237&     4.8&   4.3& 0.2$\pm$ 0.1&     0.4$\pm$    0.1& HMXB & Be \\
                 QSO B0241+62 & 41.242& 62.461&    27.5&   2.0& 0.9$\pm$ 0.1&     1.2$\pm$    0.2& Sy1 & - \\
                    4U 0253+4 & 43.602& 41.575&    12.1&   2.6& 0.7$\pm$ 0.1&     0.6$\pm$    0.1& Cluster & - \\
                {\bf AC O401} & 44.739& 13.587&     8.1&   3.2& 0.5$\pm$ 0.1&     0.9$\pm$    0.2& Cluster & - \\
               1E 0304.8+4045 & 47.052& 40.950&    39.6&   1.9& 2.2$\pm$ 0.3&     0.5$\pm$    0.1& RSCVn & - \\
         {\bf QSO J0311-7651} & 47.955&-76.856&     6.7&   3.5& 0.2$\pm$ 0.1&     0.6$\pm$    0.2& Sy1 & - \\
              RX J0317.9-4414 & 49.469&-44.233&    13.8&   2.5& 0.5$\pm$ 0.1&     0.24$\pm$    0.03& Cluster & - \\
                     NGC 1275 & 49.949& 41.516&   231.9&   1.6& 13.1$\pm$ 1.4&     7.1$\pm$    0.7& Cluster & - \\
                    H 0324+28 & 51.627& 28.693&    16.1&   2.4& 1.1$\pm$ 0.2&     0.5$\pm$    0.1& RSCVn & - \\
                      PLX 728 & 52.809& 43.886&    23.0&   2.1& 1.2$\pm$ 0.2&     2.2$\pm$    0.3& CV & IP \\
                     HD 22468 & 54.202&  0.583&    16.5&   2.4& 1.0$\pm$ 0.2&     0.4$\pm$    0.1& RSCVn & - \\
             {\bf EQ 0340-54} & 55.702&-53.614&    10.7&   2.8& 0.4$\pm$ 0.1&     -              & Cluster & - \\
        {\bf IGR J03532-6829} & 58.233&-68.511&    19.6&   2.2& 0.7$\pm$ 0.1&     0.6$\pm$    0.1& BLLac & - \\
                        X Per & 58.859& 31.072&   121.3&   1.7& 8.2$\pm$ 0.9&     7.4$\pm$    0.8& HMXB & Be \\
                    Abell 478 & 63.360& 10.467&    14.5&   2.5& 0.9$\pm$ 0.2&     0.5$\pm$    0.1& Cluster & - \\
               {\bf NGC 1566} & 64.972&-54.927&     9.4&   2.9& 0.3$\pm$ 0.1&     0.5$\pm$    0.1& Sy1 & - \\
     {\bf SWIFT J0426.2-5711} & 66.532&-57.209&    11.3&   2.7& 0.3$\pm$ 0.1&     0.23$\pm$    0.04& Sy1 & - \\
             {\bf Abell 3266} & 67.830&-61.426&    14.9&   2.4& 0.4$\pm$ 0.1&     0.5$\pm$    0.1& Cluster & - \\
                       3C 120 & 68.280&  5.373&    13.7&   2.5& 0.9$\pm$ 0.2&     1.5$\pm$    0.3& Sy1 & - \\
           {\bf X 04333-1315} & 68.391&-13.273&     8.7&   3.1& 0.5$\pm$ 0.1&     0.7$\pm$    0.2& Cluster & - \\
             {\bf 4U 0446+44} & 72.476& 45.062&     8.6&   3.1& 0.3$\pm$ 0.1&     0.5$\pm$    0.1& Cluster & - \\
              RX J0452.0+4932 & 73.015& 49.578&    20.7&   2.2& 0.9$\pm$ 0.1&     1.2$\pm$    0.3& Sy1 & - \\
  {\bf 1RXS J045602.3-753211} & 73.960&-75.529&    10.4&   2.8& 0.3$\pm$ 0.1&     0.3$\pm$    0.1& Sy2 & - \\
                     KMHK 414 & 74.563&-75.266&     9.9&   2.9& 0.3$\pm$ 0.1&       -            & StC & - \\
     {\bf SWIFT J0505.8-2351} & 76.434&-23.821&     7.6&   3.3& 0.4$\pm$ 0.1&     0.8$\pm$    0.2& Sy2 & - \\
                QSO B0502+675 & 76.998& 67.621&    20.2&   2.2& 0.7$\pm$ 0.1&     0.7$\pm$    0.1& BLLac & - \\
                  4U 0512-401 & 78.535&-40.059&   113.3&   1.7& 3.2$\pm$ 0.4&     3.0$\pm$    0.3& LMXB & G \\
          {\bf QSO B0513-002} & 79.032& -0.169&     9.8&   2.9& 0.6$\pm$ 0.1&     0.9$\pm$    0.2& Sy1 & - \\
            {\bf ESO 362-018} & 79.881&-32.624&     6.3&   3.6& 0.2$\pm$ 0.1&     0.2$\pm$    0.1& Sy1 & - \\
               {\bf PICTOR A} & 79.982&-45.785&     8.0&   3.2& 0.2$\pm$ 0.1&     0.2$\pm$    0.1& Sy1 & - \\
                      LMC X-2 & 80.052&-71.950&   784.8&   1.6& 23.0$\pm$ 2.3&    13.2$\pm$    1.3& LMXB & Z \\
         {\bf PKS J0522-3627} & 80.708&-36.447&     6.7&   3.5& 0.2$\pm$ 0.1&     0.4$\pm$    0.1& BLLac & - \\
         {\bf PKS J0525-6938} & 81.246&-69.651&     9.1&   3.0& 0.3$\pm$ 0.1&     0.21$\pm$    0.04& SNR & - \\
                       AB Dor & 82.169&-65.452&    19.1&   2.2& 0.5$\pm$ 0.1&     0.4$\pm$    0.1& Star & K1IV \\
                  3A 0527-329 & 82.342&-32.810&    31.1&   2.0& 1.1$\pm$ 0.2&     1.2$\pm$    0.2& CV & IP \\
        {\bf RX J0531.2-6607} & 82.777&-66.102&     8.5&   3.1& 0.2$\pm$ 0.1&     0.9$\pm$    0.2& HMXB & Be \\
                      LMC X-4 & 83.170&-66.376&   121.3&   1.7& 3.4$\pm$ 0.4&    10.3$\pm$    1.1& HMXB & SG \\
                         Crab & 83.616& 21.999& 19790.1&   1.6& 1000.0$\pm$ 100.1&  1000.0$\pm$  100.2& SNR & - \\
        {\bf IGR J05346-5759} & 83.725&-58.029&    19.4&   2.2& 0.5$\pm$ 0.1&     0.5$\pm$    0.17& CV & IP \\
             AXS J053514-0523 & 83.828& -5.423&    13.8&   2.5& 0.8$\pm$ 0.1&     0.8$\pm$    0.1& unid & - \\
                      LMC X-3 & 84.732&-64.085&   819.8&   1.6& 21.9$\pm$ 2.2&     3.5$\pm$    0.4& HMXB & - \\
                      LMC X-1 & 84.957&-69.736&   694.1&   1.6& 20.1$\pm$ 2.0&     3.6$\pm$    0.4& HMXB & SG \\
               PSR B0540-69.3 & 85.030&-69.351&    36.9&   1.9& 1.1$\pm$ 0.1&     1.2$\pm$    0.2& XRB & - \\
                       BY Cam & 85.689& 60.848&    15.8&   2.4& 0.6$\pm$ 0.1&     0.8$\pm$    0.1& CV & P \\
                 {\bf TX Col} & 85.820&-40.987&     5.1&   4.1& 0.15$\pm$ 0.04&     0.3$\pm$    0.1& CV & IP \\
                QSO B0549-322 & 87.706&-32.267&    11.3&   2.7& 0.4$\pm$ 0.1&     0.3$\pm$    0.1& BLLac & - \\
                     NGC 2110 & 88.047& -7.456&    11.3&   2.7& 0.6$\pm$ 0.1&     1.0$\pm$    0.2& Sy2 & - \\
        1RXS J055229.5+592842 & 88.118& 59.481&     6.2&   3.7& 0.3$\pm$ 0.1&     0.3$\pm$    0.1& Sy1 & - \\
                   4U 0558+46 & 88.747& 46.428&    28.5&   2.0& 1.1$\pm$ 0.2&     0.6$\pm$    0.1& Sy1 & - \\
        {\bf RX J0558.0+5353} & 89.456& 53.872&     7.7&   3.3& 0.3$\pm$ 0.1&     0.4$\pm$    0.1& CV & IP \\
            {\bf 4U 0557-385} & 89.494&-38.317&    16.1&   2.4& 0.5$\pm$ 0.1&     0.8$\pm$    0.1& Sy1 & (1.5) \\
          {\bf QSO B0558-504} & 89.915&-50.431&    12.7&   2.6& 0.3$\pm$ 0.1&     0.25$\pm$    0.03& Sy1 & - \\
  {\bf 1RXS J061133.8-814917} & 92.956&-81.821&    13.2&   2.6& 0.3$\pm$ 0.1&     0.6$\pm$    0.1& unid & - \\
               {\bf 4C 70.05} & 93.896& 71.038&     5.9&   3.8& 0.2$\pm$ 0.1&     0.5$\pm$    0.1& Sy2 & - \\
                  4U 0614+091 & 94.282&  9.125&   556.4&   1.6& 35.8$\pm$ 3.7&    25.3$\pm$    2.6& LMXB & B \\
             {\bf LEDA 75721} & 95.841&-64.595&     6.1&   3.7& 0.26$\pm$ 0.04&     0.6$\pm$    0.2& Cluster & - \\
        {\bf IGR J06253+7334} & 96.243& 73.587&    10.4&   2.8& 0.3$\pm$ 0.1&     0.6$\pm$    0.1& CV & IP \\
          {\bf 1ES 0625-53.6} & 96.567&-53.693&     5.6&   3.9& 0.12$\pm$ 0.04&     0.11$\pm$    0.03& Cluster & - \\
         {\bf QSO J0635-7516} & 98.944&-75.241&     5.8&   3.8& 0.15$\pm$ 0.04&     0.3$\pm$    0.1& Sy1 & - \\
  {\bf 1RXS J063847.2-535818} & 99.714&-53.962&     6.1&   3.7& 0.26$\pm$ 0.04&     0.3$\pm$    0.1& Cluster & - \\
          {\bf 1ES 0644-54.1} &101.405&-54.209&     7.6&   3.3& 0.2$\pm$ 0.1&     0.2$\pm$    0.1& Cluster & - \\
          {\bf 1ES 0646-51.5} &101.769&-51.590&     8.2&   3.2& 0.2$\pm$ 0.1&     0.2$\pm$    0.1& unid & - \\
                  {\bf Mrk 6} &103.051& 74.426&    13.2&   2.5& 0.3$\pm$ 0.1&     0.6$\pm$    0.1& Sy1 & - \\
          {\bf 1ES 0657-55.8} &104.606&-55.936&     7.4&   3.3& 0.2$\pm$ 0.1&     0.23$\pm$    0.04& Cluster & - \\
  {\bf 1RXS J070912.3-152708} &107.294&-15.454&     8.0&   3.2& 0.3$\pm$ 0.1&     0.2$\pm$    0.1& unid & - \\
            1RXS J071413.7-36 &108.568&-36.423&     5.5&   3.9& 0.2$\pm$ 0.1&     0.3$\pm$    0.1& unid & - \\
                  3A 0726-260 &112.210&-26.100&    17.2&   2.3& 0.6$\pm$ 0.1&     0.8$\pm$    0.1& HMXB & Be \\
     {\bf SWIFT J0732.5-1331} &113.154&-13.504&     5.8&   3.8& 0.3$\pm$ 0.1&     0.5$\pm$    0.2& CV & IP \\
          {\bf QSO B0738+499} &115.613& 49.846&    10.4&   2.8& 0.5$\pm$ 0.1&     0.7$\pm$    0.1& Sy1 & - \\
                   Sigma Gem. &115.803& 28.885&     9.3&   3.0& 0.7$\pm$ 0.2&     0.9$\pm$    0.2& RSCVn & - \\
  {\bf 1RXS J074616.8-161127} &116.566&-16.171&     6.4&   3.6& 0.3$\pm$ 0.1&     0.8$\pm$    0.2& unid & - \\
                   4U 0739-19 &116.875&-19.301&    23.3&   2.1& 1.0$\pm$ 0.1&     0.5$\pm$    0.1& Cluster & - \\
                 EXO 0748-676 &117.088&-67.764&   190.9&   1.6& 5.4$\pm$ 0.6&     8.8$\pm$    1.0& LMXB & B \\
              IGR J07597-3842 &119.933&-38.709&    11.8&   2.7& 0.4$\pm$ 0.1&     0.5$\pm$    0.1& Sy1 & - \\
             {\bf ES O209-12} &120.489&-49.768&     9.2&   3.0& 0.3$\pm$ 0.1&     0.3$\pm$    0.1& Sy1 & (1.5) \\
            {\bf PG 0804+761} &122.691& 76.038&    11.1&   2.7& 0.3$\pm$ 0.1&     0.4$\pm$    0.1& Sy1 & - \\
              RX J0812.4-3114 &123.140&-31.245&    26.5&   2.1& 1.0$\pm$ 0.1&     1.6$\pm$    0.2& HMXB & Be \\
        {\bf XSS J08142+6231} &123.150& 62.620&     9.2&   3.0& 0.4$\pm$ 0.1&     0.13$\pm$    0.03& CV & DN \\
      {\bf CIZA J0812.5-5714} &123.153&-57.244&     9.9&   2.9& 0.3$\pm$ 0.1&     0.3$\pm$    0.1& Cluster & - \\
                {\bf ACO 644} &124.363& -7.513&     8.6&   3.1& 0.6$\pm$ 0.1&     1.4$\pm$    0.3& Cluster & - \\
                1ES 0821-42.6 &125.823&-42.850&    15.3&   2.4& 0.5$\pm$ 0.1&     0.6$\pm$    0.1& unid & - \\
        {\bf RX J0825.2+7306} &126.300& 73.106&    14.4&   2.5& 0.4$\pm$ 0.1&     0.5$\pm$    0.1& CV & DN \\
  {\bf 1RXS J082627.1-640421} &126.493&-64.069&     7.1&   3.4& 0.2$\pm$ 0.1&     0.2$\pm$    0.1& unid. & - \\
          {\bf 1ES 0826-70.3} &126.628&-70.553&     8.5&   3.1& 0.2$\pm$ 0.1&     0.3$\pm$    0.1& unid & - \\
        {\bf RX J0832.4+3707} &128.143& 37.131&     5.0&   4.2& 0.2$\pm$ 0.1&     0.2$\pm$    0.1& Sy1 & - \\
                  Vela Pulsar &128.815&-45.183&    59.5&   1.8& 1.9$\pm$ 0.2&     2.3$\pm$    0.3& SNR & - \\
             {\bf LEDA 24297} &129.635&-35.990&     6.2&   3.7& 0.2$\pm$ 0.1&     0.4$\pm$    0.1& Sy1 & - \\
                QSO B0836+710 &130.347& 70.894&    33.4&   1.9& 1.1$\pm$ 0.1&     1.3$\pm$    0.2& QSO & - \\
                     Vela X-1 &135.519&-40.530&  1173.9&   1.6& 43.9$\pm$ 4.4&   144.2$\pm$   14.6& HMXB & SG \\
        {\bf IGR J09026-4812} &135.683&-48.211&     6.3&   3.6& 0.2$\pm$ 0.1&     0.6$\pm$    0.2& Sy1 & - \\
            {\bf X 0906-0931} &137.250& -9.689&     8.4&   3.1& 0.6$\pm$ 0.1&     0.8$\pm$    0.2& Cluster & - \\
     {\bf SWIFT J0917.2-6221} &139.020&-62.313&    13.2&   2.6& 0.4$\pm$ 0.1&     0.6$\pm$    0.1& Sy1 & - \\
                  4U 0918-549 &140.114&-55.210&   228.4&   1.6& 6.6$\pm$ 0.7&     4.8$\pm$    0.5& LMXB & B \\
                  3A 0921-630 &140.640&-63.299&    49.4&   1.8& 1.5$\pm$ 0.2&     1.3$\pm$    0.2& LMXB & D \\
              RX J0923.7+2254 &140.915& 22.896&     5.9&   3.8& 0.4$\pm$ 0.1&     0.6$\pm$    0.2& Sy1 & - \\
                   X 0922-314 &141.084&-31.700&    89.3&   1.7& 4.4$\pm$ 0.5&     2.4$\pm$    0.3& unid & - \\
{\bf SDSS J092514.05+522112.7}&141.309& 52.354&     9.5&   2.9& 0.4$\pm$ 0.1&     0.4$\pm$    0.19& QSO & - \\
             {\bf LEDA 97526} &141.511&-84.348&     6.8&   3.5& 0.14$\pm$ 0.04&     0.24$\pm$    0.04& Sy2 & - \\
                     NGC 2992 &146.396&-14.329&     7.0&   3.4& 0.5$\pm$ 0.1&     0.9$\pm$    0.2& Sy2 & - \\
                   4U 0945-30 &146.902&-30.933&    42.1&   1.9& 2.1$\pm$ 0.3&     2.0$\pm$    0.3& Sy2 & - \\
                     NGC 3031 &148.968& 69.063&    18.0&   2.3& 0.6$\pm$ 0.1&     0.3$\pm$    0.1& Sy1 & - \\
                         M 82 &148.969& 69.671&    23.0&   2.1& 0.8$\pm$ 0.1&     0.5$\pm$    0.1& Galaxy & - \\
                 GRO J1008-57 &152.451&-58.307&    16.8&   2.3& 0.5$\pm$ 0.1&     1.1$\pm$    0.2& HMXB & Be \\
        {\bf IGR J10101-5654} &152.548&-56.928&     7.3&   3.4& 0.2$\pm$ 0.1&     0.22$\pm$    0.04& HMXB & Be \\
  {\bf 1RXS J101015.9-311909} &152.571&-31.364&     8.3&   3.1& 0.4$\pm$ 0.1&     0.9$\pm$    0.2& BLLac & - \\
     {\bf SWIFT J1010.1-5747} &152.788&-57.821&     9.9&   2.9& 0.3$\pm$ 0.1&     0.7$\pm$    0.1& CV & Symb \\
               {\bf NGC 3227} &155.875& 19.862&     8.1&   3.2& 0.6$\pm$ 0.1&     1.2$\pm$    0.3& Sy1 & - \\
          {\bf QSO B1029-140} &157.964&-14.282&     5.9&   3.8& 0.3$\pm$ 0.1&     0.5$\pm$    0.1& QSO & - \\
                   4U 1036-56 &159.397&-56.791&    16.4&   2.4& 0.5$\pm$ 0.1&     1.0$\pm$    0.2& HMXB & Be \\
     {\bf SWIFT J1038.8-4942} &159.698&-49.777&     5.8&   3.8& 0.2$\pm$ 0.1&     0.14$\pm$    0.04& Sy1 & - \\
        {\bf IGR J10404-4625} &160.064&-46.431&     6.8&   3.5& 0.2$\pm$ 0.1&      -              & Sy2 & - \\
              IGR J10448-5945 &161.251&-59.700&    91.2&   1.7& 3.0$\pm$ 0.3&     2.7$\pm$    0.3& Star & EtaCar \\
  {\bf 1RXS J110337.7-232931} &165.894&-23.495&    10.2&   2.8& 0.4$\pm$ 0.1&     0.4$\pm$    0.1& BLLac & - \\
                      Mrk 421 &166.126& 38.203&   142.1&   1.6& 5.3$\pm$ 0.6&     3.5$\pm$    0.4& BLLac & - \\
                     NGC 3516 &166.706& 72.582&    34.5&   1.9& 1.0$\pm$ 0.1&     1.2$\pm$    0.2& Sy1 & - \\
                XTE J1118+480 &169.580& 48.064&    60.3&   1.8& 2.6$\pm$ 0.3&     2.5$\pm$    0.3& LMXB & M \\
                      Cen X-3 &170.318&-60.615&  1207.2&   1.6& 42.7$\pm$ 4.3&    92.9$\pm$    9.4& HMXB & Be \\
                  1H 1121-591 &171.140&-59.259&    19.6&   2.2& 0.7$\pm$ 0.1&     0.32$\pm$    0.05& SNR & - \\
        {\bf IGR J11366-6002} &174.202&-60.090&     5.7&   3.9& 0.2$\pm$ 0.1&     0.5$\pm$    0.2& AGN? & - \\
                     NGC 3783 &174.765&-37.754&    31.4&   2.0& 1.4$\pm$ 0.2&     1.4$\pm$    0.2& Sy1 & - \\
                    HD 101379 &174.834&-65.403&    65.0&   1.8& 2.3$\pm$ 0.3&     1.8$\pm$    0.2& Star & - \\
                 V* V1033 Cen &175.327&-64.183&    13.5&   2.5& 0.5$\pm$ 0.1&     0.6$\pm$    0.1& CV & P \\
           SWIFT J1142.7+7149 &175.901& 71.702&    16.6&   2.3& 0.5$\pm$ 0.1&     0.6$\pm$    0.1& CV & IP \\
              IGR J11435-6109 &176.009&-61.123&    14.9&   2.4& 0.6$\pm$ 0.1&     0.9$\pm$    0.2& HMXB & Be \\
        {\bf RX J1145.2+7940} &176.294& 79.687&    13.2&   2.5& 0.3$\pm$ 0.1&     0.6$\pm$    0.1& Sy1 & - \\
             {\bf H 1143-182} &176.432&-18.460&    10.5&   2.8& 0.5$\pm$ 0.1&     1.0$\pm$    0.2& Sy1 & - \\
               1E 1145.1-6141 &176.847&-61.961&   148.9&   1.6& 5.3$\pm$ 0.6&    12.4$\pm$    1.3& HMXB & SG \\
               2E 1145.5-6155 &176.993&-62.196&    32.1&   2.0& 1.2$\pm$ 0.2&     1.6$\pm$    0.2& HMXB & Be \\
        {\bf RX J1155.4-5641} &178.851&-56.722&     7.6&   3.3& 0.3$\pm$ 0.1&     0.5$\pm$    0.1& CV & DN \\
               {\bf NGC 3998} &179.496& 55.451&     7.4&   3.3& 0.3$\pm$ 0.1&     0.4$\pm$    0.1& AGN & liner \\
        {\bf IGR J12026-5349} &180.723&-53.839&     5.8&   3.8& 0.2$\pm$ 0.1&     0.4$\pm$    0.1& Sy2 & - \\
                     NGC 4151 &182.649& 39.407&    88.1&   1.7& 3.3$\pm$ 0.4&     6.5$\pm$    0.7& Sy1 & - \\
  {\bf 1RXS J121222.7-580118} &183.130&-58.010&     7.7&   3.3& 0.3$\pm$ 0.1&     0.5$\pm$    0.1& unid & - \\
               EXMS B1210-645 &183.308&-64.887&    50.0&   1.8& 1.8$\pm$ 0.2&     2.2$\pm$    0.3& unid & - \\
       {\bf SAX J1221.7+7526} &185.529& 75.385&     8.8&   3.0& 0.2$\pm$ 0.1&     0.4$\pm$    0.1& Sy2 & -\\
                     GX 301-2 &186.707&-62.776&   562.6&   1.6& 20.3$\pm$ 2.1&   121.5$\pm$   12.4& HMXB & SG \\
                       3C 273 &187.272&  2.041&    19.7&   2.2& 2.4$\pm$ 0.4&     2.4$\pm$    0.4& BLLac & - \\
        {\bf IGR J12349-6434} &188.771&-64.557&    15.4&   2.4& 0.6$\pm$ 0.1&     1.8$\pm$    0.3& CV & Symb \\
        {\bf IGR J12415-5750} &190.348&-57.826&     9.1&   3.0& 0.4$\pm$ 0.1&     0.7$\pm$    0.2& Sy2 & (1.5) \\
            {\bf 1H 1249-637} &190.673&-63.049&    19.6&   2.2& 0.7$\pm$ 0.1&     0.9$\pm$    0.1& HMXB & Be \\
                   Abell 3526 &192.207&-41.316&    17.3&   2.3& 0.9$\pm$ 0.1&     0.5$\pm$    0.1& Cluster & - \\
                   4U 1246-58 &192.389&-59.072&   167.9&   1.6& 6.8$\pm$ 0.7&     5.2$\pm$    0.6& LMXB & B \\
              RX J1252.4-2914 &193.122&-29.256&    52.6&   1.8& 2.9$\pm$ 0.4&     1.8$\pm$    0.2& CV & IP \\
                  4U 1254-690 &194.395&-69.308&   881.1&   1.6& 30.0$\pm$ 3.0&    20.7$\pm$    2.1& LMXB & B \\
                     ACO 1656 &194.928& 27.965&    23.7&   2.1& 1.3$\pm$ 0.2&     1.0$\pm$    0.1& Cluster & - \\
        {\bf IGR J13020-6359} &195.473&-63.974&    12.2&   2.6& 0.4$\pm$ 0.1&     1.0$\pm$    0.2& HMXB & Be? \\
  {\bf 1RXS J131651.8-715537} &199.261&-71.913&    12.2&   2.6& 0.4$\pm$ 0.1&     0.5$\pm$    0.1& unid & - \\
                     NGC 5128 &201.359&-43.008&   199.2&   1.6& 10.7$\pm$ 1.1&    19.5$\pm$    2.1& Sy2 & - \\
                   4U 1323-62 &201.605&-62.127&   110.7&   1.7& 4.9$\pm$ 0.5&     7.6$\pm$    0.8& LMXB & B \\
       {\bf RXC J1327.9-3130} &202.040&-31.468&     7.6&   3.3& 0.4$\pm$ 0.1&     0.9$\pm$    0.2& Cluster & - \\
                MCG-06-30-015 &203.983&-34.289&    17.6&   2.3& 1.0$\pm$ 0.2&     1.4$\pm$    0.2& Sy1 & - \\
                 1ES 1344-326 &206.888&-32.838&    24.1&   2.1& 1.4$\pm$ 0.2&     1.0$\pm$    0.2& Cluster & - \\
                   Abell 1795 &207.226& 26.612&    17.9&   2.3& 0.9$\pm$ 0.1&     0.7$\pm$    0.1& RGal & - \\
                  3A 1346-301 &207.334&-30.306&    56.0&   1.8& 3.4$\pm$ 0.4&     3.9$\pm$    0.5& Sy1 & - \\
                {\bf Mrk 279} &208.264& 69.321&    20.2&   2.2& 0.8$\pm$ 0.1&     0.4$\pm$    0.1& Sy1 & - \\
                  GS 1354-645 &209.495&-64.733&    41.2&   1.9& 1.8$\pm$ 0.2&     2.6$\pm$    0.3& LMXB & BHC \\
  {\bf 1RXS J140041.2-632623} &210.220&-63.445&    13.2&   2.6& 0.6$\pm$ 0.1&     0.8$\pm$    0.2& unid & - \\
          {\bf Circinus Gal.} &213.293&-65.342&    12.1&   2.6& 0.5$\pm$ 0.1&     1.4$\pm$    0.3& Sy2 & - \\
                     NGC 5506 &213.302& -3.233&    37.7&   1.9& 2.4$\pm$ 0.3&     3.9$\pm$    0.5& Sy2 & - \\
  {\bf 1RXS J141656.0-120053} &214.200&-11.981&     5.9&   3.8& 0.4$\pm$ 0.1&     1.1$\pm$    0.3& unid & - \\
           {\bf 2E 1415+2556} &214.495& 25.704&     8.0&   3.2& 0.4$\pm$ 0.1&     0.6$\pm$    0.1& BLLac & - \\
               {\bf NGC 5548} &214.515& 25.125&    16.8&   2.3& 0.8$\pm$ 0.1&     0.7$\pm$    0.1& Sy1 & (1.5) \\
                   H 1417-624 &215.302&-62.706&    44.5&   1.9& 2.2$\pm$ 0.3&     6.4$\pm$    0.8& HMXB & Be \\
     {\bf NSC J142605+374853} &216.535& 37.806&     7.6&   3.3& 0.3$\pm$ 0.1&     1.0$\pm$    0.2& Cluster & - \\
             {\bf H 1426+428} &217.151& 42.659&    11.3&   2.7& 0.5$\pm$ 0.1&     1.0$\pm$    0.2& BLLac & - \\
             SAX J1428.6-5422 &217.214&-54.390&    37.7&   1.9& 2.4$\pm$ 0.3&     3.0$\pm$    0.4& unid & - \\
         {\bf WFC J1452-5708} &223.913&-57.146&    13.7&   2.5& 1.0$\pm$ 0.2&     1.3$\pm$    0.2& unid & - \\
        {\bf IGR J15094-6649} &227.455&-66.831&    14.2&   2.5& 0.7$\pm$ 0.1&     0.9$\pm$    0.2& CV & IP \\
                    Abell 202 &227.755&  5.741&    13.0&   2.6& 0.9$\pm$ 0.2&     0.6$\pm$    0.1& Cluster & - \\
                 PSR B1509-58 &228.494&-59.156&    26.8&   2.0& 1.9$\pm$ 0.3&     1.9$\pm$    0.3& XP & - \\
               {\bf ACO 2052} &229.198&  6.984&     5.7&   3.9& 0.4$\pm$ 0.1&     1.1$\pm$    0.3& Cluster & - \\
          {\bf QSO B1517+656} &229.381& 65.417&    10.5&   2.8& 0.4$\pm$ 0.1&     0.25$\pm$    0.03& BLLac & - \\
                      Cir X-1 &230.152&-57.171& 10565.0&   1.6& 933.1$\pm$ 93.5&   502.3$\pm$   50.4& LMXB & A \\
       {\bf RXC J1539.5-8335} &234.934&-83.585&    19.7&   2.2& 0.4$\pm$ 0.1&     -                 & unid & - \\
                  4U 1538-522 &235.601&-52.367&   123.6&   1.7& 10.4$\pm$ 1.1&    26.9$\pm$    3.0& HMXB & SG \\
                XTE J1543-568 &235.982&-56.776&    23.5&   2.1& 1.9$\pm$ 0.3&     5.4$\pm$    0.8& HMXB & Be \\
                  4U 1543-624 &236.967&-62.544&   489.5&   1.6& 30.7$\pm$ 3.1&    22.5$\pm$    2.3& LMXB & - \\
                XTE J1550-564 &237.706&-56.484&   929.2&   1.6& 76.3$\pm$ 7.7&    23.6$\pm$    2.4& LMXB & M \\
            {\bf PG 1553+113} &238.939& 11.168&     6.7&   3.5& 0.5$\pm$ 0.1&     1.5$\pm$    0.4& QSO & - \\
         {\bf PKS J1557-7913} &239.212&-79.228&     9.4&   2.9& 0.2$\pm$ 0.1&     0.3$\pm$    0.1& Sy2 & - \\
                   Abell 2142 &239.617& 27.247&    22.7&   2.1& 1.1$\pm$ 0.2&     1.2$\pm$    0.2& Cluster & - \\
                  1H 1556-605 &240.273&-60.730&   255.2&   1.6& 17.2$\pm$ 1.8&    11.5$\pm$    1.2& LMXB & - \\
        {\bf XSS J16019-7548} &240.369&-75.756&    19.8&   2.2& 0.5$\pm$ 0.1&     0.4$\pm$    0.1& Cluster & - \\
                  4U 1608-522 &243.152&-52.415&   159.2&   1.6& 13.8$\pm$ 1.5&    11.3$\pm$    1.2& LMXB & A \\
          {\bf MCG+14-08-004} &244.876& 81.035&     5.8&   3.8& 0.18$\pm$ 0.04&     0.11$\pm$    0.02& Galaxy & - \\
\renewcommand{\thefootnote}{\arabic{footnote}}~\setcounter{footnote}{1} SCO X-1\footnote{\scriptsize{data containig SCOX-1 were too noisy to be added in the total mosaic map}}  &  244.975  & -15.624 & - & -&  LMXB & Z \\
        {\bf RX J1625.5+8529} &246.740& 85.492&     6.4&   3.6& 0.16$\pm$ 0.03&     0.4$\pm$    0.1& Sy1 & - \\
                  4U 1624-490 &247.001&-49.171&   547.9&   1.6& 51.8$\pm$ 5.3&    50.5$\pm$    5.2& LMXB & D \\
                   Abell 2199 &247.190& 39.550&    27.9&   2.0& 0.9$\pm$ 0.1&     0.6$\pm$    0.1& Cluster & - \\
              AX J1631.9-4752 &248.004&-47.883&    10.1&   2.9& 1.0$\pm$ 0.2&     3.1$\pm$    0.6& HMXB & - \\
               {\bf ACO 3628} &247.808&-75.135&     9.4&   2.9& 0.3$\pm$ 0.1&     0.3$\pm$    0.1& Cluster & - \\
                   4U 1626-67 &248.066&-67.456&   128.5&   1.7& 5.8$\pm$ 0.6&    12.0$\pm$    1.3& LMXB & - \\
                   4U 1630-47 &248.529&-47.394&   115.5&   1.7& 11.7$\pm$ 1.3&     7.9$\pm$    0.9& LMXB & BHC \\
               {\bf ACO 2218} &248.954& 66.207&     6.3&   3.6& 0.2$\pm$ 0.1&     0.4$\pm$    0.1& Cluster & - \\
              IGR J16377-6423 &249.582&-64.358&    26.4&   2.1& 1.4$\pm$ 0.2&     1.2$\pm$    0.2& Cluster & - \\
         {\bf WFC J1640-2001} &250.068&-20.018&    32.9&   2.0& 4.7$\pm$ 0.6&     3.5$\pm$    0.5& unid & - \\
               {\bf ACO 2219} &250.076& 46.713&    10.0&   2.9& 0.3$\pm$ 0.1&    -               & Cluster & - \\
                  4U 1636-536 &250.207&-53.752&  2550.6&   1.6& 213.4$\pm$ 21.4&   150.5$\pm$   15.1& LMXB & - \\
                     GX 340+0 &251.435&-45.623&  4296.7&   1.6& 452.9$\pm$ 45.4&   352.4$\pm$   35.4& LMXB & - \\
         {\bf WFC J1649-1818} &252.307&-18.302&    28.7&   2.0& 3.9$\pm$ 0.5&     5.6$\pm$    0.8& unid & - \\
                      Mrk 501 &253.466& 39.778&   128.8&   1.7& 4.0$\pm$ 0.4&     3.2$\pm$    0.4& BLLac & - \\
                 GRO J1655-40 &253.479&-39.836&  2286.6&   1.6& 232.7$\pm$ 23.4&    65.1$\pm$    6.6& LMXB & - \\
        {\bf IGR J16558-5203} &254.032&-52.062&    13.6&   2.5& 1.1$\pm$ 0.2&     1.0$\pm$    0.2& Sy1 & - \\
                      Her X-1 &254.467& 35.337&   245.5&   1.6& 8.7$\pm$ 0.9&    22.8$\pm$    2.4& LMXB & - \\
                 OAO 1657-415 &255.186&-41.673&    42.0&   1.9& 4.1$\pm$ 0.5&    15.4$\pm$    2.0& HMXB & SG \\
                   H 1658-298 &255.494&-29.955&   135.6&   1.7& 12.5$\pm$ 1.3&    11.4$\pm$    1.22& LMXB & - \\
                     GX 339-4 &255.665&-48.786&   182.8&   1.6& 16.1$\pm$ 1.7&     8.1$\pm$    0.9& LMXB & - \\
               {\bf ACO 2244} &255.690& 34.073&     8.0&   3.2& 0.3$\pm$ 0.1&     0.2$\pm$    0.1& Cluster & - \\
             EXSS 1706.6+7842 &255.992& 78.620&    29.7&   2.0& 0.7$\pm$ 0.1&     1.0$\pm$    0.1& Cluster? & - \\
                  4U 1700-377 &255.998&-37.859&   439.9&   1.6& 43.5$\pm$ 4.5&   106.6$\pm$   11.0& HMXB & SG \\
                     GX 349+2 &256.429&-36.416&  6965.4&   1.6& 711.3$\pm$ 71.3&   597.8$\pm$   60.0& LMXB & - \\
                  4U 1702-429 &256.542&-43.054&   378.3&   1.6& 35.2$\pm$ 3.6&    32.3$\pm$    3.3& LMXB & - \\
                   4U 1700+24 &256.640& 23.959&    16.0&   2.4& 0.9$\pm$ 0.1&     1.0$\pm$    0.2& LMXB & - \\
                  4U 1705-440 &257.219&-44.130&  1154.2&   1.6& 104.3$\pm$ 10.5&    82.5$\pm$    8.3& LMXB & - \\
              RX J1709.5-2639 &257.347&-26.686&    24.6&   2.1& 2.2$\pm$ 0.3&     1.8$\pm$    0.3& LMXB & - \\
         {\bf EXMS B1707-375} &257.753&-37.618&     9.5&   2.9& 0.9$\pm$ 0.2&     0.7$\pm$    0.1& unid & - \\
             SAX J1711.6-3808 &257.928&-38.109&    35.4&   1.9& 3.2$\pm$ 0.4&     3.9$\pm$    0.5& LMXB & - \\
         {\bf WFC J1712-2639} &258.057&-26.660&    26.6&   2.0& 2.3$\pm$ 0.3&     0.7$\pm$    0.1& unid & - \\
                   4U 1708-40 &258.073&-40.851&   183.0&   1.6& 16.6$\pm$ 1.7&     9.5$\pm$    1.0& LMXB & - \\
            {\bf OPH CLUSTER} &258.095&-23.378&    28.4&   2.0& 2.8$\pm$ 0.4&     2.5$\pm$    0.3& Cluster & - \\
             SAX J1712.6-3739 &258.136&-37.664&    48.2&   1.8& 4.5$\pm$ 0.5&     3.9$\pm$    0.5& LMXB & - \\
     {\bf NSC J171252+640307} &258.189& 64.052&     9.8&   2.9& 0.3$\pm$ 0.1&     0.21$\pm$    0.04& Cluster & - \\
                  2S 1711-339 &258.578&-34.029&    55.7&   1.8& 4.9$\pm$ 0.6&     4.3$\pm$    0.5& LMXB & - \\
                XTE J1716-389 &258.983&-38.859&   132.2&   1.7& 11.6$\pm$ 1.3&     8.6$\pm$    1.0& HMXB & SG \\
        {\bf RX J1719.2+4858} &259.838& 48.989&     7.5&   3.3& 0.26$\pm$ 0.04&     0.5$\pm$    0.1& Sy1 & - \\
 {\bf FIRST J172201.9+431523} &260.465& 43.256&     6.1&   3.7& 0.23$\pm$ 0.04&    -0.2$\pm$   -0.1& Sy1 & - \\
                XTE J1723-376 &260.896&-37.687&    73.0&   1.7& 6.2$\pm$ 0.7&     4.6$\pm$    0.5& LMXB & - \\
        {\bf RX J1723.8+8553} &261.309& 85.881&     8.9&   3.0& 0.23$\pm$ 0.04&     0.3$\pm$    0.1& Cluster & - \\
                  4U 1724-307 &261.866&-30.791&   357.6&   1.6& 27.3$\pm$ 2.8&    36.6$\pm$    3.8& LMXB & G \\
          {\bf QSO B1727+502} &262.063& 50.235&    11.2&   2.7& 0.3$\pm$ 0.1&     0.21$\pm$    0.04& BLLac & - \\
                    HD 159023 &262.165& 59.056&    14.0&   2.5& 0.3$\pm$ 0.1&     -               & Star & - \\
                       GX 9+9 &262.934&-16.951&  2277.3&   1.6& 236.4$\pm$ 23.8&   174.8$\pm$   17.6& LMXB & - \\
                     GX 354-0 &262.984&-33.860&  1189.2&   1.6& 91.5$\pm$ 9.2&   133.5$\pm$   13.5& LMXB & - \\
                       GX 1+4 &263.020&-24.719&   135.8&   1.7& 11.2$\pm$ 1.2&    49.1$\pm$    5.3& LMXB & - \\
                1ES 1734+74.2 &263.241& 74.238&    10.4&   2.8& 0.3$\pm$ 0.1&     0.3$\pm$    0.1& RSCVn & - \\
                  4U 1730-335 &263.351&-33.389&   286.7&   1.6& 21.7$\pm$ 2.2&    22.2$\pm$    2.3& LMXB & G \\
                  KS 1731-260 &263.573&-26.088&  1251.6&   1.6& 97.6$\pm$ 9.9&    79.0$\pm$    8.0& LMXB & - \\
                 SLX 1735-269 &264.574&-26.990&    94.1&   1.7& 6.8$\pm$ 0.8&     8.5$\pm$    1.0& LMXB & - \\
         {\bf WFC J1738-3336} &264.613&-33.601&    35.1&   1.9& 2.6$\pm$ 0.3&     1.2$\pm$    0.2& unid & - \\
                  4U 1735-444 &264.724&-44.467&  1926.8&   1.6& 151.4$\pm$ 15.2&   134.1$\pm$   13.5& LMXB & - \\
  {\bf 1RXS J174018.3-205103} &265.111&-20.840&    27.9&   2.0& 2.5$\pm$ 0.3&     1.9$\pm$    0.3& unid & - \\
                XTE J1739-278 &265.687&-27.749&   103.7&   1.7& 7.3$\pm$ 0.8&     1.7$\pm$    0.2& LMXB & - \\
{\bf 2MASX J17431735+6250207} &265.811& 62.822&     6.7&   3.5& 0.25$\pm$ 0.04&     0.3$\pm$    0.1& Sy2 & - \\
               1E 1740.7-2942 &265.956&-29.728&   200.5&   1.6& 13.5$\pm$ 1.4&    56.4$\pm$    5.9& LMXB & - \\
      {\bf 1RXS J1744.1+3259} &266.061& 32.987&     8.5&   3.1& 0.3$\pm$ 0.1&     0.5$\pm$    0.1& unid & - \\
                 GRO J1744-28 &266.156&-28.756&   256.6&   1.6& 17.0$\pm$ 1.8&    54.4$\pm$    5.7& LMXB & - \\
                  1A 1742-294 &266.534&-29.520&   355.0&   1.6& 23.7$\pm$ 2.4&    45.2$\pm$    4.7& LMXB & - \\
         {\bf 1E 1743.1-2843} &266.573&-28.733&    35.1&   1.9& 2.3$\pm$ 0.3&     7.4$\pm$    1.0& LMXB & - \\
             SAX J1747.0-2853 &266.719&-28.869&   138.0&   1.7& 9.1$\pm$ 1.0&    10.8$\pm$    1.2& LMXB & - \\
                 SLX 1744-299 &266.879&-30.029&   254.2&   1.6& 17.3$\pm$ 1.8&    24.7$\pm$    2.6& LMXB & - \\
                       GX 3+1 &266.969&-26.572&  4335.2&   1.6& 318.5$\pm$ 31.9&   285.6$\pm$   28.7& LMXB & - \\
                 EXO 1745-248 &267.028&-24.781&   442.4&   1.6& 33.9$\pm$ 3.5&    39.9$\pm$    4.1& LMXB & G \\
                  4U 1745-203 &267.225&-20.357&    43.2&   1.9& 3.8$\pm$ 0.5&     5.0$\pm$    0.6& LMXB & G \\
                  4U 1746-370 &267.528&-37.075&   458.4&   1.6& 34.6$\pm$ 3.5&    38.2$\pm$    3.9& LMXB & G \\
             SAX J1750.8-2900 &267.565&-29.035&    36.8&   1.9& 2.5$\pm$ 0.3&     4.2$\pm$    0.5& LMXB & - \\
                 GRS 1747-312 &267.684&-31.285&    76.1&   1.7& 5.3$\pm$ 0.6&     7.1$\pm$    0.8& LMXB & G \\
                       GX 5-1 &270.307&-25.086& 11222.9&   1.6& 991.4$\pm$ 99.3&   761.9$\pm$   76.4& LMXB & - \\
                 GRS 1758-258 &270.308&-25.731&   192.6&   1.6& 15.8$\pm$ 1.7&    23.4$\pm$    2.5& LMXB & - \\
                       GX 9+1 &270.383&-20.520&  7216.2&   1.6& 725.5$\pm$ 72.8&   562.7$\pm$   56.4& LMXB & - \\
             SAX J1808.4-3658 &272.149&-36.980&    24.8&   2.1& 2.0$\pm$ 0.3&     1.6$\pm$    0.2& LMXB & - \\
                      GX 13+1 &273.625&-17.141&  3095.4&   1.6& 349.4$\pm$ 35.1&   183.5$\pm$   18.4& LMXB & - \\
                   4U 1812-12 &273.779&-12.101&    90.6&   1.7& 10.4$\pm$ 1.2&    12.9$\pm$    1.4& LMXB & - \\
                      GX 17+2 &274.019&-14.015&  5025.8&   1.6& 609.8$\pm$ 61.1&   493.3$\pm$   49.5& LMXB & - \\
                       AM Her &274.061& 49.869&   114.2&   1.7& 2.9$\pm$ 0.3&     4.0$\pm$    0.4& CV & P \\
\renewcommand{\thefootnote}{\arabic{footnote}}~\setcounter{footnote}{2} SAX J1818-1703/WFC J1818-1658\footnote{\scriptsize{see Section~\ref{saxj_wfc}}}&-16.970&    32.5&   2.0& 3.7$\pm$ 0.5&     3.0$\pm$    0.4& unid & - \\
  {\bf 1RXS J182129.0-131641} &275.331&-13.308&    15.5&   2.4& 1.7$\pm$ 0.3&     1.0$\pm$    0.2& unid & - \\
               QSO J1821+6420 &275.504& 64.342&    24.5&   2.1& 0.6$\pm$ 0.1&     0.6$\pm$    0.1& Sy1 & - \\
                  4U 1820-303 &275.927&-30.374&  2772.3&   1.6& 236.6$\pm$ 23.8&   217.3$\pm$   21.9& LMXB & G \\
                  4U 1822-000 &276.333& -0.029&   309.2&   1.6& 24.9$\pm$ 2.6&    16.3$\pm$    1.7& LMXB & - \\
                  4U 1822-371 &276.477&-37.113&   182.9&   1.6& 15.5$\pm$ 1.6&    33.9$\pm$    3.6& LMXB & - \\
                   GS 1826-24 &277.362&-23.784&   312.9&   1.6& 31.2$\pm$ 3.2&    39.6$\pm$    4.1& LMXB & - \\
         {\bf SNR 021.5-00.9} &278.382&-10.559&    12.4&   2.6& 1.2$\pm$ 0.2&     3.0$\pm$    0.5& SNR & - \\
                       3C 382 &278.761& 32.680&    27.9&   2.0& 1.1$\pm$ 0.1&     0.8$\pm$    0.1& Sy1 & - \\
                RX JB1832-330 &278.926&-33.020&    49.0&   1.8& 4.2$\pm$ 0.5&     4.5$\pm$    0.5& LMXB & G \\
            {\bf 1H 1828-593} &279.222&-59.400&    12.2&   2.6& 0.6$\pm$ 0.1&     0.5$\pm$    0.1& Sy2 & - \\
             {\bf ESO 103-35} &279.553&-65.434&    13.9&   2.5& 0.6$\pm$ 0.1&     1.7$\pm$    0.3& Sy2 & - \\
                      Ser X-1 &279.978&  5.052&  2934.4&   1.6& 227.1$\pm$ 22.8&   161.0$\pm$   16.2& LMXB & - \\
                     3C 390.3 &280.628& 79.777&    33.3&   1.9& 0.8$\pm$ 0.1&     1.0$\pm$    0.1& Sy1 & - \\
       {\bf RX J184452-62215} &281.201&-62.383&    11.9&   2.7& 0.5$\pm$ 0.1&     1.1$\pm$    0.2& Sy1 & - \\
               Ginga 1843+009 &281.413&  0.823&    60.7&   1.8& 4.7$\pm$ 0.6&    18.1$\pm$    2.1& HMXB & Be \\
             {\bf 4U 1916-79} &281.750&-78.532&    11.6&   2.7& 0.3$\pm$ 0.1&     0.7$\pm$    0.1& Sy1 & - \\
              IGR J18483-0311 &282.073& -3.211&    11.1&   2.7& 0.9$\pm$ 0.2&     2.2$\pm$    0.4& HMXB & - \\
                  3A 1845-024 &282.088& -2.461&    17.1&   2.3& 1.4$\pm$ 0.2&     5.1$\pm$    0.8& HMXB & Be \\
        {\bf IGR J18485-0047} &282.144& -0.731&     9.0&   3.0& 0.7$\pm$ 0.2&     0.9$\pm$    0.2& unid & - \\
                  4U 1850-087 &283.270& -8.694&    28.8&   2.0& 2.8$\pm$ 0.4&     2.7$\pm$    0.4& LMXB & G \\
         {\bf 1E 1846.5-7857} &283.722&-78.907&     7.4&   3.3& 0.26$\pm$ 0.04&     0.6$\pm$    0.1& Sy1 & - \\
                XTE J1855-026 &283.909& -2.622&     7.9&   3.2& 0.7$\pm$ 0.2&     3.3$\pm$    0.7& HMXB & - \\
                XTE J1856+053 &284.169&  5.321&    34.7&   1.9& 2.7$\pm$ 0.4&     0.9$\pm$    0.1& LMXB & - \\
         {\bf 1E 1849.2-7832} &284.362&-78.463&     8.2&   3.1& 0.22$\pm$ 0.04&     0.4$\pm$    0.1& Sy1 & - \\
                XTE J1859+226 &284.673& 22.653&   210.8&   1.6& 10.1$\pm$ 1.1&     2.2$\pm$    0.2& LMXB & - \\
          {\bf XTE J1901+014} &285.412&  1.449&     9.0&   3.0& 0.7$\pm$ 0.2&     0.7$\pm$    0.1& XRB & - \\
         {\bf WFC J1907+1305} &286.943& 13.088&    12.7&   2.6& 0.9$\pm$ 0.2&     1.1$\pm$    0.1& unid & - \\
          {\bf 1ES 1907+52.3} &287.084& 52.425&     7.9&   3.2& 0.2$\pm$ 0.1&     0.14$\pm$    0.02& RSCV & - \\
                  4U 1907+097 &287.413&  9.830&    68.3&   1.7& 5.3$\pm$ 0.6&    13.2$\pm$    1.5& HMXB & SG \\
         {\bf WFC J1909+1246} &287.497& 12.769&    14.7&   2.4& 1.0$\pm$ 0.2&     0.9$\pm$    0.2& unid & - \\
                   4U 1909+07 &287.723&  7.608&    38.7&   1.9& 3.1$\pm$ 0.4&     8.0$\pm$    1.0& HMXB & - \\
        {\bf AX J1911.0+0906} &287.785&  9.088&    16.4&   2.4& 1.3$\pm$ 0.2&     0.6$\pm$    0.1& unid & - \\
                      Aql X-1 &287.831&  0.597&   366.1&   1.6& 31.2$\pm$ 3.2&    19.7$\pm$    2.0& LMXB & - \\
                       SS 433 &287.961&  4.988&    27.1&   2.0& 2.3$\pm$ 0.3&     3.2$\pm$    0.4& HMXB & SG \\
              IGR J19140+0951 &288.510&  9.864&    12.1&   2.6& 0.9$\pm$ 0.2&     1.8$\pm$    0.3& HMXB & SG \\
                 GRS 1915+105 &288.818& 10.948&  7057.8&   1.6& 596.3$\pm$ 59.7&   482.5$\pm$   48.4& LMXB & - \\
                  4U 1916-053 &289.684& -5.234&   107.6&   1.7& 9.4$\pm$ 1.0&     8.9$\pm$    1.0& LMXB & - \\
                   4U 1919+44 &290.292& 43.960&    33.7&   1.9& 1.2$\pm$ 0.2&     1.3$\pm$    0.2& Cluster & - \\
                   ESO 141-55 &290.339&-58.680&    16.5&   2.4& 0.8$\pm$ 0.1&     1.1$\pm$    0.2& Sy1 & - \\
                    V* CH Cyg &291.129& 50.237&     7.3&   3.4& 0.2$\pm$ 0.1&     0.3$\pm$    0.1& CV & Symb \\
        {\bf RX J1927.3+6533} &291.857& 65.556&    10.2&   2.8& 0.3$\pm$ 0.1&     0.4$\pm$    0.1& Sy1 & - \\
         {\bf WFC J1929+1720} &292.250& 17.338&    17.3&   2.3& 1.0$\pm$ 0.2&     1.1$\pm$    0.2& unid & - \\
         {\bf WFC J1933+1408} &293.252& 14.141&    17.2&   2.3& 1.1$\pm$ 0.2&     1.2$\pm$    0.2& unid & - \\
         {\bf WFC J1935+2053} &293.996& 20.894&    15.7&   2.4& 0.8$\pm$ 0.1&     1.1$\pm$    0.2& unid & - \\
                XTE J1946+274 &296.403& 27.329&   115.2&   1.7& 5.3$\pm$ 0.6&    12.5$\pm$    1.4& HMXB & Be \\
  {\bf 1RXS J194708.6-762335} &296.734&-76.403&     8.4&   3.1& 0.2$\pm$ 0.1&     0.3$\pm$    0.1& unid & - \\
                  KS 1947+300 &297.390& 30.172&    45.7&   1.8& 2.0$\pm$ 0.3&     4.6$\pm$    0.6& HMXB & Be \\
        {\bf XSS J19303-7950} &297.486&-79.764&     8.2&   3.2& 0.21$\pm$ 0.04&     0.4$\pm$    0.1& unid. & - \\
                  4U 1954+319 &298.947& 32.096&   128.2&   1.7& 5.9$\pm$ 0.6&    13.1$\pm$    1.4& HMXB & - \\
                      Cyg X-1 &299.578& 35.220&  6690.7&   1.6& 337.1$\pm$ 33.78&   370.4$\pm$   37.2& HMXB & M \\
                  4U 1957+115 &299.850& 11.702&   379.8&   1.6& 28.5$\pm$ 2.9&     6.8$\pm$    0.7& LMXB & - \\
                        Cyg A &299.859& 40.713&    30.6&   2.0& 1.3$\pm$ 0.2&     1.9$\pm$    0.3& QSO & - \\
                QSO B1959+650 &300.046& 65.159&    62.7&   1.8& 1.7$\pm$ 0.2&     0.9$\pm$    0.1& BLLac & - \\
           SWIFT J2009.0-6103 &302.189&-61.110&     9.1&   3.0& 0.4$\pm$ 0.1&     0.4$\pm$    0.1& Sy1 & - \\
               QSO J2009-4849 &302.374&-48.827&    77.2&   1.7& 3.2$\pm$ 0.4&     2.2$\pm$    0.3& BLLac & - \\
                1E S2008-57.0 &303.078&-56.849&    12.6&   2.6& 0.5$\pm$ 0.1&     0.6$\pm$    0.1& Cluster & - \\
              RX J2012.6+3809 &303.216& 38.161&    11.4&   2.7& 0.5$\pm$ 0.1&     0.6$\pm$    0.1& LMXB & - \\
        {\bf RX J2014.4+6123} &303.620& 61.401&     6.9&   3.5& 0.2$\pm$ 0.1&     0.2$\pm$    0.1& unid. & - \\
                 EXO 2030+375 &308.061& 37.627&    25.7&   2.1& 1.3$\pm$ 0.2&     2.6$\pm$    0.4& HMXB & Be \\
                      Cyg X-3 &308.136& 40.948&  3438.0&   1.6& 165.7$\pm$ 16.6&   167.0$\pm$   16.8& HMXB & - \\
             {\bf LEDA 64989} &308.649&-30.632&     6.8&   3.5& 0.4$\pm$ 0.1&     0.4$\pm$    0.1& Sy1 & - \\
               {\bf 4C 74.26} &310.674& 75.121&    28.3&   2.0& 0.7$\pm$ 0.1&     1.1$\pm$    0.2& QSO & - \\
                      Mrk 509 &311.060&-10.733&    16.2&   2.4& 1.0$\pm$ 0.2&     0.9$\pm$    0.1& Sy1 & - \\
  {\bf 1RXS J204937.4-800800} &312.323&-80.128&    10.0&   2.9& 0.2$\pm$ 0.1&     0.3$\pm$    0.1& RSCV & - \\
  {\bf 1RXS J205528.2-002123} &313.877& -0.365&     8.7&   3.1& 0.6$\pm$ 0.1&     1.2$\pm$    0.3& BLLac & - \\
       {\bf IGR J202569+4940} &314.167& 49.680&    18.9&   2.2& 0.6$\pm$ 0.1&     0.8$\pm$    0.1& AGN? & - \\
           {\bf GRO J2058+42} &314.734& 41.782&    11.0&   2.7& 0.5$\pm$ 0.1&     1.1$\pm$    0.2& HMXB & Be \\
             SAX J2103.5+4545 &315.920& 45.736&    56.5&   1.8& 2.2$\pm$ 0.3&     4.5$\pm$    0.5& HMXB & Be \\
  {\bf 1RXS J210604.3+614322} &316.553& 61.755&     5.2&   4.1& 0.28$\pm$ 0.04&     0.13$\pm$    0.04& unid. & - \\
        {\bf XSS J21128+8216} &318.529& 82.092&    20.3&   2.2& 0.4$\pm$ 0.1&     0.8$\pm$    0.1& Sy1 & - \\
              IGR J21247+5058 &321.153& 50.981&    23.1&   2.1& 0.8$\pm$ 0.1&     1.2$\pm$    0.2& Sy1 & - \\
     {\bf SWIFT J2127.4+5654} &321.934& 56.947&    13.5&   2.5& 0.4$\pm$ 0.1&     0.5$\pm$    0.1& Sy1 & - \\
                   4U 2129+12 &322.512& 12.176&   180.9&   1.6& 12.5$\pm$ 1.3&     7.6$\pm$    0.8& LMXB & G \\
        1RXS J213202.3-334255 &323.023&-33.719&    11.2&   2.7& 0.5$\pm$ 0.1&     0.7$\pm$    0.1& Sy1 & - \\
        {\bf IGR J21335+5105} &323.426& 51.122&    12.4&   2.6& 0.4$\pm$ 0.1&     0.5$\pm$    0.1& CV & IP \\
         {\bf QSO J2136-6224} &324.129&-62.394&     6.6&   3.5& 0.3$\pm$ 0.1&     0.6$\pm$    0.2& Sy1 & - \\
  {\bf 1RXS J213833.0+320507} &324.671& 32.096&     7.7&   3.3& 0.5$\pm$ 0.1&     0.15$\pm$    0.02& Sy1 & - \\
                       SS Cyg &325.660& 43.595&    89.0&   1.7& 3.9$\pm$ 0.4&     4.2$\pm$    0.5& CV & DN \\
                      Cyg X-2 &326.174& 38.325&  8269.0&   1.6& 522.2$\pm$ 52.3&   317.4$\pm$   31.8& LMXB & - \\
         {\bf PKS J2151-3027} &327.975&-30.477&     6.6&   3.6& 0.3$\pm$ 0.1&     0.3$\pm$    0.1& QSO & Blazar \\
         {\bf PKS J2157.6941} &329.265&-69.692&     6.6&   3.5& 0.2$\pm$ 0.1&     0.3$\pm$    0.1& Sy1 & - \\
               PKS J2158-3013 &329.742&-30.242&    23.5&   2.1& 1.2$\pm$ 0.2&     0.8$\pm$    0.1& BLLac & - \\
  {\bf 1RXS J220157.8-595648} &330.463&-59.949&     8.2&   3.2& 0.3$\pm$ 0.1&     0.6$\pm$    0.1& Cluster & - \\
               {\bf NGC 7172} &330.491&-31.865&     6.0&   3.8& 0.3$\pm$ 0.1&     0.8$\pm$    0.2& Sy2 & - \\
                  4U 2206+543 &332.019& 54.512&   116.2&   1.7& 3.8$\pm$ 0.4&     5.4$\pm$    0.6& HMXB & Be \\
               2E 2206.1-4724 &332.338&-47.189&    12.0&   2.6& 0.5$\pm$ 0.1&     0.6$\pm$    0.1& Sy1 & - \\
              RX J2214.0+1242 &333.529& 12.689&    10.6&   2.8& 0.7$\pm$ 0.1&     1.3$\pm$    0.3& CV & DN \\
              {\bf V* FO Aqr} &334.474& -8.350&    15.2&   2.4& 1.0$\pm$ 0.2&     1.8$\pm$    0.3& CV & IP \\
  {\bf 1RXS J223355.0-843406} &338.685&-84.588&     7.9&   3.2& 0.24$\pm$ 0.04&     0.5$\pm$    0.1& unid. & - \\
               {\bf NGC 7314} &338.937&-26.016&     7.5&   3.3& 0.4$\pm$ 0.1&     0.15$\pm$    0.02& Sy1 & - \\
                   4U 2238+60 &339.812& 61.261&     9.0&   3.0& 0.3$\pm$ 0.1&     0.8$\pm$    0.2& HMXB & Be \\
                {\bf Ark 564} &340.661& 29.752&    14.7&   2.4& 0.7$\pm$ 0.1&     0.21$\pm$    0.03& Sy1 & - \\
                       IM peg &343.302& 16.829&    14.1&   2.5& 0.7$\pm$ 0.1&     0.5$\pm$    0.1& RSCVn & - \\
               {\bf 3C 454.3} &343.484& 16.153&     6.2&   3.7& 0.3$\pm$ 0.1&     0.9$\pm$    0.2& Star & - \\
          {\bf QSO B2251-178} &343.535&-17.573&    11.5&   2.7& 0.8$\pm$ 0.2&     1.1$\pm$    0.2& Sy1 & - \\
      {\bf 1AXG J225518-0310} &343.846& -3.189&    22.4&   2.1& 1.3$\pm$ 0.2&     1.4$\pm$    0.2& CV & IP \\
         {\bf 2E 2259.0+5836} &345.299& 58.881&    12.4&   2.6& 0.4$\pm$ 0.1&     0.3$\pm$    0.1& HMXB & - \\
  {\bf 1RXS J230238.1+713649} &345.654& 71.613&     8.2&   3.2& 0.2$\pm$ 0.1&         -          & unid & - \\
               {\bf NGC 7469} &345.791&  8.874&    14.7&   2.5& 0.9$\pm$ 0.2&     1.4$\pm$    0.2& Sy1 & - \\
               {\bf NGC 7582} &349.600&-42.351&     9.7&   2.9& 0.4$\pm$ 0.1&     0.9$\pm$    0.2& Sy2 & - \\
                        Cas A &350.885& 58.831&  1271.2&   1.6& 43.2$\pm$ 4.4&    11.8$\pm$    1.2& SNR & - \\
                       II Peg &358.788& 28.639&     7.3&   3.4& 0.3$\pm$ 0.1&     0.7$\pm$    0.2& RSCVn & - \\
         {\bf PKS J2357-3445} &359.235&-34.769&     9.9&   2.9& 0.5$\pm$ 0.1&     0.6$\pm$    0.1& Cluster & - \\
                   H 2356-309 &359.810&-30.616&    10.4&   2.8& 0.5$\pm$ 0.1&     0.8$\pm$    0.2& BLLac & - \\
  {\bf 1RXS J000053.0-783038} &359.987&-78.523&     6.1&   3.7& 0.14$\pm$ 0.04&     0.7$\pm$    0.2& unid & - \\

\footnotetext[1]{\scriptsize{a name in bold face indicates a new detection with respect to \citet{Verrecchia}.}}
\footnotetext[2]{\scriptsize{error circle radius extrapolated from Figure~\ref{errore} with the systematic error included. }}
\footnotetext[3]{\scriptsize{average flux estimation in mCrab (see Section~\ref{curve} for details).}} 
\footnotetext[4]{\scriptsize{
Type classifications: 
AGN=Active galactic nuclei; 
BLLac=BL Lac object;  
Cluster=Cluster of galaxies; 
CV=Cataclysmic variable; 
HMXB=High-mass X-ray binary; 
LMXB=Low-mass X-ray binary; 
QSO = Quasar; 
RGal=Radio Galaxy;  
RSCVn= RS Canum Venaticorum variable; 
SNR=Supernova remnant; 
Sy1=Seyfert 1 galaxy;  
Sy2=Seyfert 2 galaxy;
Unid=Unidentified source;
StC=Star Cluster
XRB=Galactic X-ray binary; 
XP=X-ray pulsar; 
}}
\footnotetext[5]{\scriptsize{Sub Type classifications:
A=Atoll-type source (neutron Star);
B=Burster (neutron star); 
Be=B-type emission-line star;  
BHC=Black hole candidate;
D=Dipping; 
DN= Dawrf Nova;
G=Globular Cluster X-ray source;
IP=Intermediate Polar;
M=microquasar
P=Polar;
Symb=Symbiotic star;
SG=Supergiant;
Z=Z-type source; }}
\end{longtable}
\end{center}
\normalsize
\renewcommand{\thefootnote}{\arabic{footnote}}
\clearpage
\end{landscape} 
\clearpage

\begin{table}[t!]
 \begin{center}
\scriptsize
\caption{{\it WFC new source candidates}}
 \label{tab:newsources}
 \begin{tabular}{lcccccc}
 \hline
 \hline
  Name & R.A. & Dec & sigma & err. rad. & exposure &  possible counterpart\tablenotemark{1}\\
 \hline
  -& Deg. & Deg. & - & arcmin & s$\times$10$^{6}$  &- \\ 
 \hline
 \hline

{\bf WFC J1452-5708}    & 223.913  & -57.146  & 13.7  & 2.5 & 7.5   & 2MASS J14553991-5710008 \\  
{\bf WFC J1640-2001}    & 250.068  & -20.018  & 32.9  & 2.0 & 4.6   & - \\   
{\bf WFC J1649-1818}    & 252.307  & -18.302  & 28.7  & 2.0 & 4.6   & - \\          
{\bf WFC J1712-2639}    & 258.057  & -26.660  & 26.6  & 2.0 & 5.2   & - \\       
{\bf WFC J1738-3336}    & 264.613  & -33.601  & 35.1  & 1.9 & 5.6   & 2MASS J17381495-3335373 \\             
{\bf WFC J1907+1305}    & 286.943  & 13.088  & 12.7   & 2.6 & 4.0   & - \\        
{\bf WFC J1935+2053}    & 293.996  & 20.894  & 15.7   & 2.4 & 4.9   & 2MASS J19360455+2055306 \\ 
{\bf WFC J1909+1246}    & 287.497  & 12.769  & 14.7   & 2.4 & 3.9   & - \\         
{\bf WFC J1929+1720}    & 292.250  & 17.338  & 17.3   & 2.3 & 4.8   & - \\       
{\bf WFC J1933+1408}    & 293.252  & 14.141  & 17.2   & 2.3 & 4.5   & - \\      

\hline
 \hline
\end{tabular}
\end{center}
\end{table} 


\begin{acknowledgements}
The authors would like to thank the Rome section of the IASF-INAF institute for the financial support.
Special thanks also go to the ASI Data center staff and in particular to Francesco Verrecchia for the useful BeppoSAX WFC material made available on line through the ADSC web site.
The authors acknowledge financial contribution from the agreement ASI-INAF  I/009/10/0 and I/033/10/0. 
FC also thanks the Physics Departiment of {\it Universit\'a degli studi Roma 3} for the kind hospitality.
\end{acknowledgements}

\end{document}